\documentclass[11pt,a4paper]{article}
\pdfoutput=1
\usepackage{jinstpub}
\usepackage{graphicx}
\usepackage{textgreek}
\usepackage{amsmath}
\usepackage{amssymb}
\usepackage{placeins}
\usepackage{multirow}
\usepackage{circuitikzgit}
\usepackage{subfigure}
\usepackage{capt-of}
\usepackage{booktabs}
\usepackage{parskip}
\usepackage{varwidth}
\usepackage{soul}

\title{Investigation of ASIC-based signal readout electronics for LEGEND-1000}

\author[a,b,1]{F.~Edzards \note{Corresponding authors},}
\author[c,b,a,1]{M.~Willers,}
\author[d]{A.~Alborini,}
\author[d]{L.~Bombelli,}
\author[a]{D.~Fink,}
\author[e,f,g]{M.~P.~Green,}
\author[h]{M.~Laubenstein,}
\author[a,b]{S.~Mertens,}
\author[i,f]{G.~Othman,}
\author[g]{D.~C.~Radford,}
\author[b]{S.~Sch\"onert}
\author[j]{and G.~Zuzel}

\affiliation[a]{Max Planck Institute for Physics, F\"ohringer Ring 6, 80805 Munich, Germany}
\affiliation[b]{Technical University of Munich, Arcisstrasse 21, 80333 Munich, Germany}
\affiliation[c]{Lawrence Berkeley National Laboratory, 1 Cyclotron Rd, CA 94720, US}
\affiliation[d]{XGLab SRL, Bruker Nano Analytics, Via Conte Rosso 23, 20134 Milano, Italy}
\affiliation[e]{North Carolina State University, Raleigh, NC 27695, US}
\affiliation[f]{Triangle Universities Nuclear Laboratory, Durham, NC, US}
\affiliation[g]{Oak Ridge National Laboratory, 1 Bethel Valley Rd, TN 37830, US}
\affiliation[h]{Laboratori Nazionali del Gran Sasso, Via G.~Acitelli 22, 67100 Assergi, Italy}
\affiliation[i]{University of North Carolina-Chapel Hill, 120 E.~Cameron Ave, NC 27599, US}
\affiliation[j]{M.~Smoluchowski Institute of Physics, Jagiellonian University, 30-348 Krak$\acute o$w, Poland}

\emailAdd{edzards@mpp.mpg.de}
\emailAdd{mwillers@mpp.mpg.de}

\abstract{
LEGEND, the Large Enriched Germanium Experiment for Neutrinoless $\text{\textbeta\textbeta}$ Decay, is a ton-scale experimental program to search for neutrinoless double beta ($0\text{\textnu\textbeta\textbeta}$) decay in the isotope $^{76}$Ge with an unprecedented sensitivity. Building on the success of the low-background $^{76}$Ge-based GERDA and \textsc{Majorana Demonstrator} experiments, the LEGEND collaboration is targeting a signal discovery sensitivity beyond $10^{28}\,$yr on the decay half-life with approximately $10\,\text{t}\cdot\text{yr}$ of exposure. Signal readout electronics in close proximity to the detectors plays a major role in maximizing the experiment's discovery sensitivity by reducing electronic noise and improving pulse shape analysis capabilities for the rejection of backgrounds. However, the proximity also poses unique challenges for the radiopurity of the electronics. Application-specific integrated circuit (ASIC) technology allows the implementation of the entire charge sensitive amplifier (CSA) into a single low-mass chip while improving the electronic noise and reducing the power consumption. In this work, we investigated the properties and electronic performance of a commercially available ASIC CSA, the XGLab CUBE preamplifier, together with a p-type point contact high-purity germanium detector. We show that low noise levels and excellent energy resolutions can be obtained with this readout. Moreover, we demonstrate the viability of pulse shape discrimination techniques for reducing background events.
}

\keywords{LEGEND, signal readout electronics, ASIC, high-purity germanium detectors}

\notoc
\begin{document}
\maketitle
\section{Introduction}
The observation of neutrinoless double beta ($0\text{\textnu\textbeta\textbeta}$) decay would have major implications on our understanding of the origin of matter in our universe. The decay violates lepton number conservation by two units and is the most sensitive way to obtain information on whether neutrinos are Majorana particles, i.e.~their own antiparticles. Moreover, together with measurements from cosmology and direct neutrino mass measurements, it will provide information on the absolute neutrino mass scale and ordering~\cite{dolinski2019, giuliani2020}. One of the most promising technologies in the search for $0\text{\textnu\textbeta\textbeta}$ decay are high-purity germanium (HPGe) detectors. Germanium detectors are intrinsically pure, can be enriched readily to about 88\% and above in the double beta decaying isotope $^{76}$Ge, and provide an excellent energy resolution of about 0.1\%~FWHM in the signal region of interest at $Q_{\text{\textbeta\textbeta}}=2039\,$keV. The LEGEND collaboration employs a phased approach to realize the ultimate goal of a ton-scale $0\text{\textnu\textbeta\textbeta}$ decay search with HPGe detectors~\cite{abgrall2017, myslik2019, abgrall2018}.

The main requirements for readout electronics in $0\text{\textnu\textbeta\textbeta}$ decay experiments employing $^{76}$Ge are good energy resolution (low electronic noise), good pulse shape discrimination (PSD) capabilities (separation of signal events from background events), and a high radiopurity (low background). To obtain a high energy resolution and good PSD capabilities, the readout electronics needs to be placed as close as possible to the detector to minimize stray input capacitance~\cite{abgrall2014}. However, this is in conflict with the radiopurity requirement, i.e.~any component in close proximity to the detector contributes more to the radioactive background~\cite{cuesta2017}. A promising approach is to combine all relevant readout electronics components into a single low-mass, low-background chip located very close to the detector using application-specific integrated circuit (ASIC) technology. In this work, we investigated the performance of a commercially available ASIC with regard to the requirements of LEGEND.

\section{LEGEND}\label{ch:legend}
The LEGEND collaboration has been formed to pursue a ton-scale $^{76}$Ge-based $0\text{\textnu\textbeta\textbeta}$ decay experiment utilizing the best technologies from the GERDA (GERmanium Detector Array) and \mbox{\textsc{Majorana Demonstrator}} experiments, as well as contributions from other groups. GERDA and \textsc{Majorana Demonstrator} have achieved the lowest backgrounds ($4\cdot10^{-4}$\,counts/(keV\,$\cdot$\,kg\,$\cdot$\,yr), held by GERDA) and the best energy resolutions (2.5\,keV FWHM at $Q_{\text{\textbeta\textbeta}}$, held by the \textsc{Majorana Demonstrator}) of all experimental $0\text{\textnu\textbeta\textbeta}$ decay searches \cite{agostini2019, wiesinger2020, aalseth2018, alvis2019}. To achieve a signal discovery sensitivity at a half-life of $T_{1/2}^{0\text{\textnu}}>10^{28}\,$yr, LEGEND pursues a phased approach. In the first phase, LEGEND-200, up to 200\,kg of HPGe detectors will be operated in the cryogenic infrastructure previously installed by the GERDA collaboration at the Laboratori Nazionali del Gran Sasso (LNGS). The detectors will be operated in liquid argon which acts both as a cooling medium and as an active shielding. At the same time, ultra-high radiopurity materials for all internal structures and low-noise signal readout electronics will be used. The overall background is estimated to improve by a factor of more than two compared to the background achieved in the GERDA experiment to a level below $2\cdot10^{-4}\,$counts/(keV\,$\cdot$\,kg\,$\cdot$\,yr), with a targeted signal discovery sensitivity of $T_{1/2}^{0\text{\textnu}}>10^{27}\,$yr. In the final stage of LEGEND, LEGEND-1000, the collaboration plans to operate 1000\,kg of HPGe detectors for a time period of about 10\,years. This requires a completely new infrastructure and a more ambitious background goal of less than $1\cdot10^{-5}\,$counts/(keV\,$\cdot$\,kg\,$\cdot$\,yr) to reach the targeted signal discovery sensitivity on the half-life of $T_{1/2}^{0\text{\textnu}}>10^{28}\,$yr.

\section{Signal readout electronics for LEGEND}\label{ch:eletronics_legend}

\subsection{Readout electronics requirements}
The charge sensitive amplifier (CSA) should be located as close as possible to the detector. This reduces the capacitive load on the amplifier and is necessary to keep the electronic noise level of the system as low as possible. High noise levels increase the energy threshold and degrade the energy resolution thereby decreasing the experimental sensitivity. Another advantage of a close proximity of the CSA to the detector is the enhanced bandwidth of the system, i.e.~faster signal rise times, which are important for the successful application of pulse shape analysis (PSA) techniques to reject background events. The noise and rise time requirements are in conflict with the radiopurity requirements. The components close to the detectors contribute to the radioactive background budget and decrease the experimental sensitivity. It is therefore desirable to have as little material as possible close to the detectors. Consequently, the material mass and volume of the CSA needs to be very small. In addition, the selected components must be very radiopure. In conclusion, one has to find a good compromise between low noise levels and fast rise times on the one hand, and low radioactivity of the components close to the detectors on the other.

\subsection{Readout electronics for LEGEND}
\begin{figure}[!h]
\begin{center}
\begin{circuitikz}[scale=0.8, transform shape] 
\draw
(-2.6,-1.3) node[ground]{} to [ioosource, l=$Q\delta(t)$, -] (-2.6,0)
(-0.95,-1.3) node[ground]{} to [capacitor, l=$C_{\text{det}}$, -*] (-0.95,0) 
(-2.6,0) -- (2.5,0)

\pgfextra{\ctikzset{tripoles/nmos/width=.45,
                    tripoles/nmos/height=.45}}
(2.8,0) node[nmos](nmosA){} node[right=1.2mm]{JFET}(2.8,0)
(2.5,0) -- (nmosA.G)

\pgfextra{\ctikzset{tripoles/plain amp/width=.88,
                    tripoles/plain amp/height=.88}}
(7,0) node[plain amp] (plainamp) {} node[] {\hspace{-2mm}A}[]
(nmosA.D) to (plainamp.-)
(nmosA.S) to (plainamp.+)
(plainamp.out) -- (9,0)
(2,0) -- (2,2.8)
to [generic, l=$R_{\text{f}}$] (4,2.8)

(2,1.3) to [capacitor, l=$C_{\text{f}}$, *-*] (4,1.3)
(4,2.8) -- (4,1.3) -- (7.7,1.3)
(7.7,0) node[circ]{} to (7.7,1.3)
(2,0) node[circ]{} to (2,-1.5) to [capacitor, l=$C_{\text{p}}$] (4,-1.5) -- (9,-1.5)
; 

\draw[-latex] (-3.0,-1.2) -- (-3.0,-0.3); 

\draw [black!30]
(0.8,-1.3) node[ground]{} to [capacitor, l=$C_{\text{par}}$, -*, color=black!30] (0.8,0) 
;
 





\draw [dashed] 
(1.5,3.5) -- (1.5,-2.6)
(4.5,3.5) -- (4.5,-2.6)
(6.0,3.5) -- (6.0,-2.6)
(8.3,3.5) -- (8.3,-2.6)
;

\draw
(-0.79,-2.4) node[] {Detector + Bias voltage}[]
(3.1,-2.4) node[] {Front-end}[]
(5.25,-2.4) node[] {Cables}[]
(7.14,-2.45) node[] {Preamplifier}[]
;
\end{circuitikz}
\end{center}
\caption{Simplified circuit of the resistive-feedback signal readout electronics that will be used in LEGEND-200. The front-end stage (close to the detectors) and the preamplifier stage (farther away) are separated by cables. Both stages will be operated at cryogenic temperatures in liquid argon. In LEGEND-1000, all relevant components will potentially be combined into a single low-mass ASIC.}
\label{graph:preamp_overview}
\end{figure}
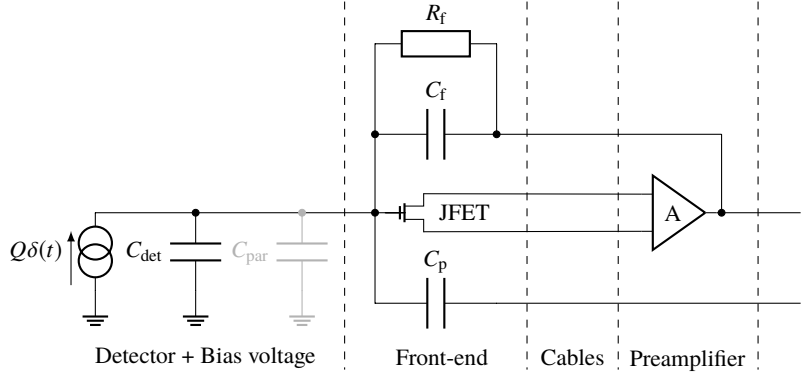

In LEGEND-200, electronic components based on previous implementations by the predecessor experiments, GERDA and \textsc{Majorana Demonstrator}, will be used. The CSA consists of two stages: A first stage very close to the detectors (several cm) is based on \textsc{Majorana Demonstrator}'s radiopure low-noise, low-mass front-end (LMFE) readout electronics \cite{abgrall2015}. The LMFE consists of a junction field-effect transistor (JFET) and an $RC$ feedback circuit, see figure~\ref{graph:preamp_overview}. A second stage farther away (${\sim}30\,$cm above the detector array) has been developed based on the preamplifier of the GERDA experiment \cite{riboldi2011}.

One of the main challenges when scaling up a germanium-based $0\text{\textnu\textbeta\textbeta}$ decay experiment is the increased number of individual detectors, resulting in an increase of instrumentation components, such as amplifiers, cables and connectors. All these components are potential background sources and hence need to be of ultra-high purity and low mass \cite{abgrall2018}. For LEGEND-1000, the baseline design is to use an ASIC-based readout scheme for the HPGe detectors. State-of-the art ASIC technology enables the integration and miniaturization of the readout electronics components into a single low-mass chip. The main advantage for $0\text{\textnu\textbeta\textbeta}$ decay searches compared to conventional amplifiers is a potentially higher per-channel radiopurity (e.g.~ideally no $RC$ components, fewer supply voltages, etc.). Furthermore, a lower electronic noise can be achieved since ASIC technology allows for a high amplification gain close to the detector before sending the analog signal over a long distance to the data acquisition system. 
\FloatBarrier
\noindent

\subsection{CUBE ASIC}
\begin{figure}[!h]
\begin{center}
\includegraphics[angle=0,width=0.75\textwidth]{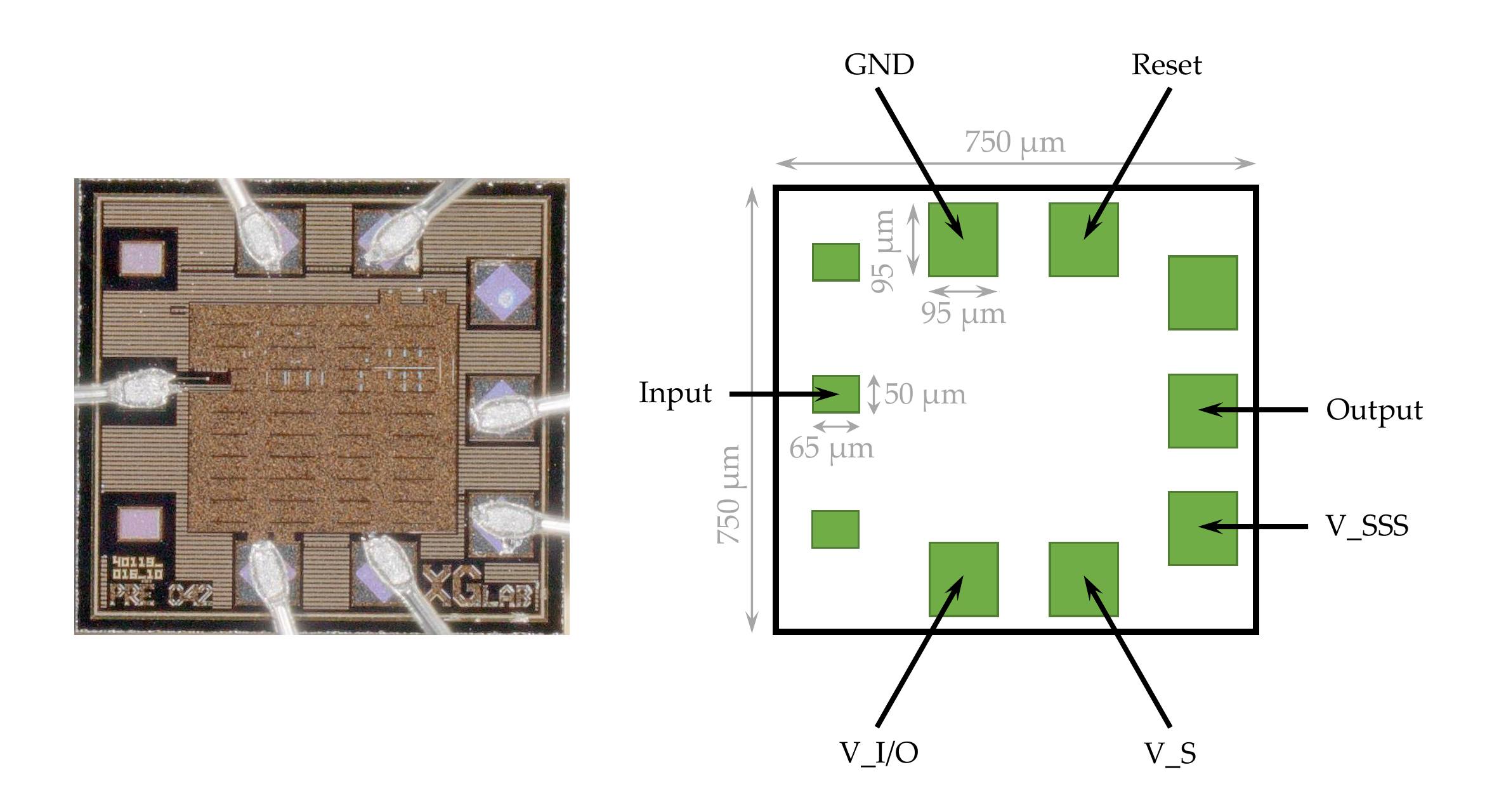}
\caption{The left figure shows a close-up of the CUBE ASIC. The chip is connected to the traces on a printed circuit board via several wire bonds. The right figure shows the dimensions of the ASIC and the bond pad assignment. The chip requires three supply voltages \texttt{V\_I/O}, \texttt{V\_S} and \texttt{V\_SSS}.}
\label{graph:cube_dimensions}
\end{center}
\end{figure}

\begin{figure}[!h]
\begin{center}
\includegraphics[angle=0,width=0.7\textwidth]{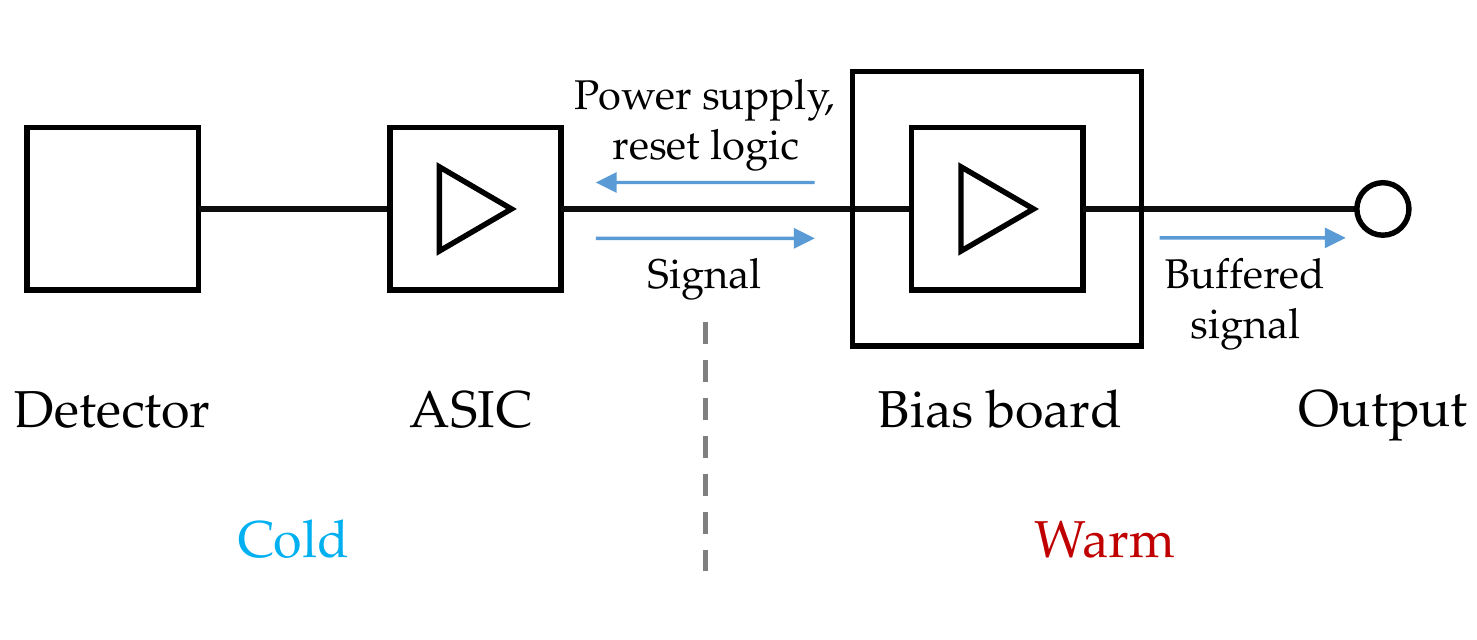}
\caption{Detection and amplification chain in our CUBE ASIC measurements. The germanium detector and the ASIC were operated in a cold environment in a vacuum cryostat. Outside of the cryostat, a biasing board provided the supply voltages, as well as reset logic for the CUBE ASIC. In addition, the board buffered the output signal with a certain gain.}
\label{graph:amp_chain}
\end{center}
\end{figure}

\begin{figure}[!h]
\begin{center}
\vspace{-0.35cm}
\includegraphics[angle=0,width=0.62\textwidth]{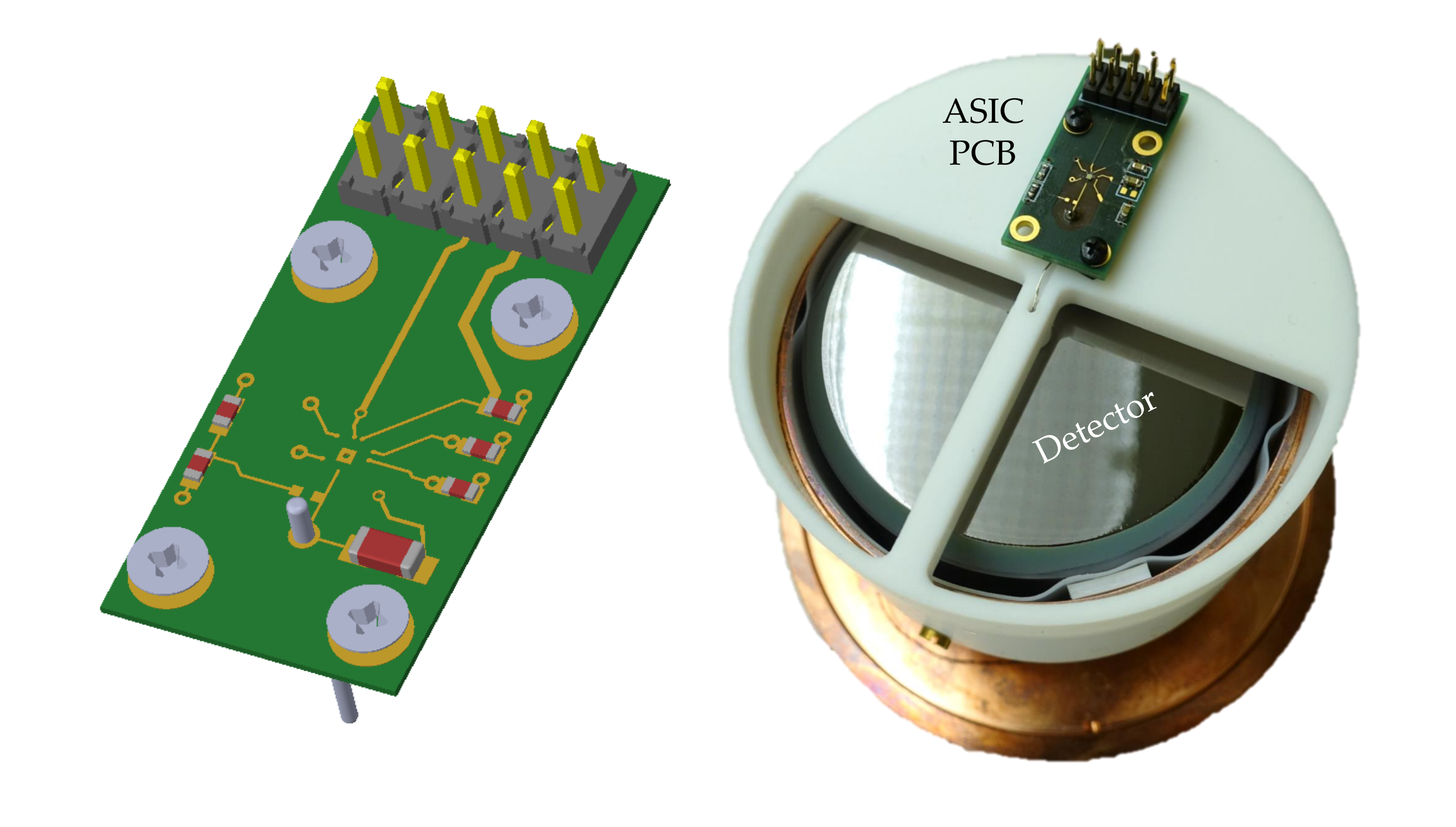}
\caption{Printed circuit board hosting the CUBE ASIC. The left figure shows a three-dimensional rendering of the board. Three bypass capacitors are used for reducing the noise generated by the supply voltages. Moreover, a voltage divider can be used to test the functionality of the ASIC with an external pulse generator. The right figure shows the ASIC board mounted on a PTFE structure above the PPC germanium detector.}
\label{graph:carrier_board}
\end{center}
\end{figure}

To test the performance of ASIC-based signal readout electronics, a p-type point contact (PPC) detector (cf.~section~\ref{ch:cube_facility}) was instrumented with a commercially available CUBE ASIC obtained from the company XGLab SRL. The CUBE ASIC is a low-noise CSA based on CMOS technology. It was initially designed for low-capacitance (several pF) silicon drift detectors \cite{bombelli2011}. In a first study of the CUBE ASIC (revision \texttt{PRE\_024}) with a low-capacitance mini PPC detector, a noise performance of 5.6\,e$^-$~RMS was obtained \cite{barton2016}. With no capacitive load at the input, the ASIC investigated in this work (revision \texttt{PRE\_042}) has a noise performance of 35.5\,e$^-$~RMS at room temperature and is optimized for operation with detectors having higher capacitances~\cite{xglab2019}. An image of the chip and an annotated illustration of its wire bonding pads is shown in figure~\ref{graph:cube_dimensions}. The ASIC measures $750\,\text{\textmu m}\times750\,\text{\textmu m}\times250\,\text{\textmu m}$ and has a mass of $0.33\,$mg. It has an input capacitance which is optimized for the operation of detectors with capacitances in the range $0.5\leq C_{\text{det}}\leq 3.0\,$pF. The chip is functional at cryogenic temperatures down to 50\,K and has a maximum power consumption of 60\,mW~\cite{xglab2019}. The internal feedback capacitance of $C_{\text{f}}=500\,\text{fF}\pm10\%$ corresponds to a dynamic range with energies larger than 10\,MeV in germanium at cryogenic temperatures.
\\\\
The CUBE ASIC requires three supply voltages provided by an external biasing board, see figure~\ref{graph:amp_chain}. Each of these supplies needs at least one external bypass capacitor to reduce the voltage supply noise. The effectiveness of these capacitors decreases with increasing distance to the ASIC. Therefore, they need to be mounted as close as possible to the preamplifier. Unfortunately, the bypass capacitors also increase the amount of radioactive material close to the detectors. Usually, they are not clean enough to fulfill the stringent radiopurity requirements \cite{abgrall2018}. For our measurements, a customized printed circuit board accommodating the CUBE ASIC was designed. Figure~\ref{graph:carrier_board} shows a drawing of the board and its mounting on top of the PPC germanium detector.
\FloatBarrier
\noindent

\begin{figure}[!h]
\begin{center}
\includegraphics[angle=0,width=0.6\textwidth]{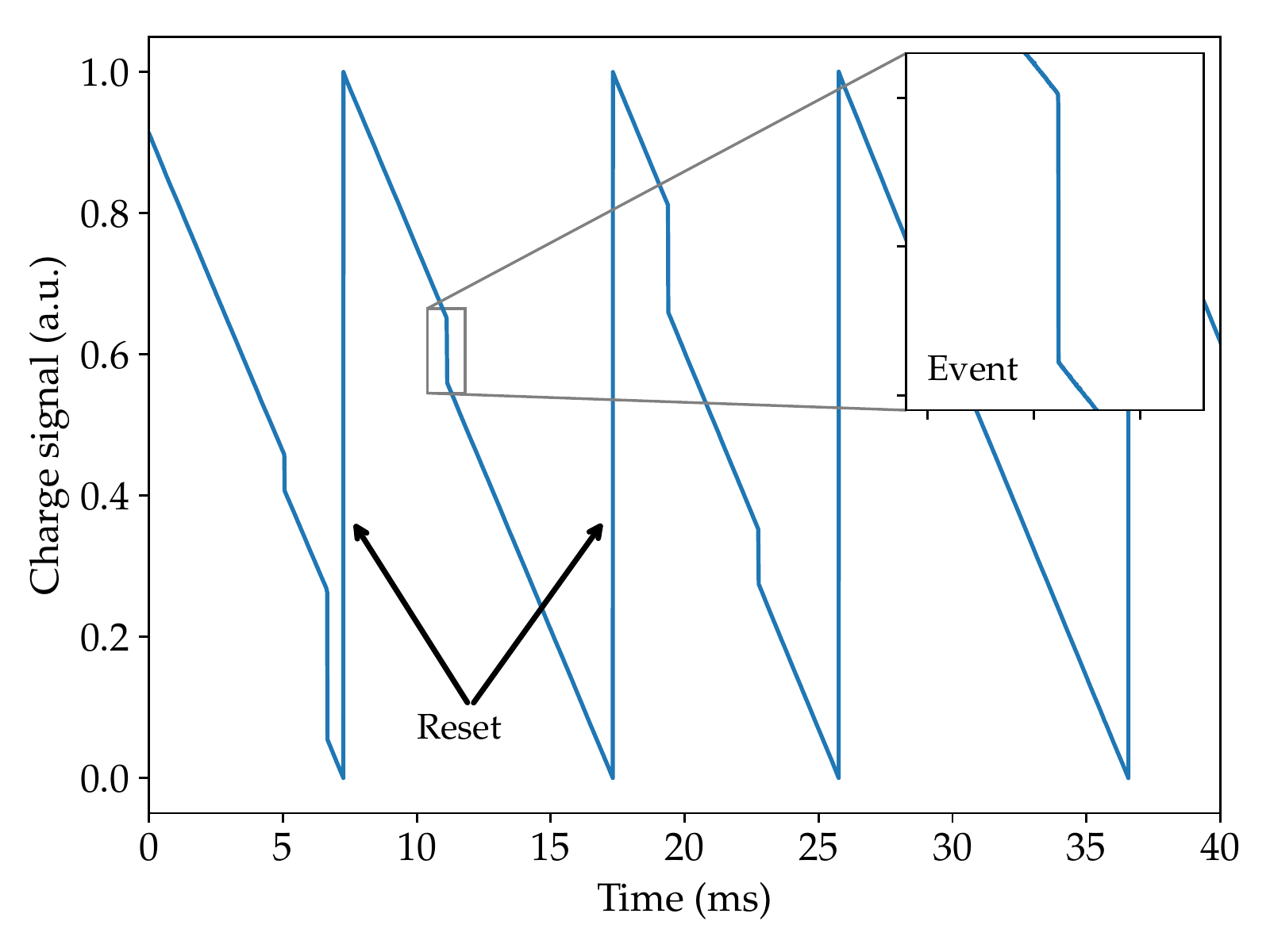}
\caption{Waveform example obtained with the CUBE ASIC operated in the pulsed reset mode. Events can be identified as steps in the linearly decreasing ramps (see inset). Since there is no feedback resistor removing the charges from the feedback capacitance, every event decreases the preamplifier output to a lower voltage. As soon as the dynamic range of the preamplifier is reached, it is reset back to the starting value by an external feedback device. The reset events can be identified as the large positive steps between the decreasing ramps.}
\label{graph:cube_waveforms}
\end{center}
\end{figure}

\FloatBarrier
\noindent
In order to avoid saturation and a reduction of the dynamic range of the readout electronics, the ASIC needs to be reset appropriately. By default, it is operated in a pulsed reset mode. In this mode, the CUBE preamplifier uses an external logic signal to control a CMOS transistor to discharge the feedback capacitor. A waveform example is shown in figure~\ref{graph:cube_waveforms}.
\FloatBarrier
\noindent

\FloatBarrier
\noindent

\section{Measurement setup}\label{ch:cube_facility}
\begin{figure}[!h]
\begin{center}
\includegraphics[angle=0,width=0.85\textwidth]{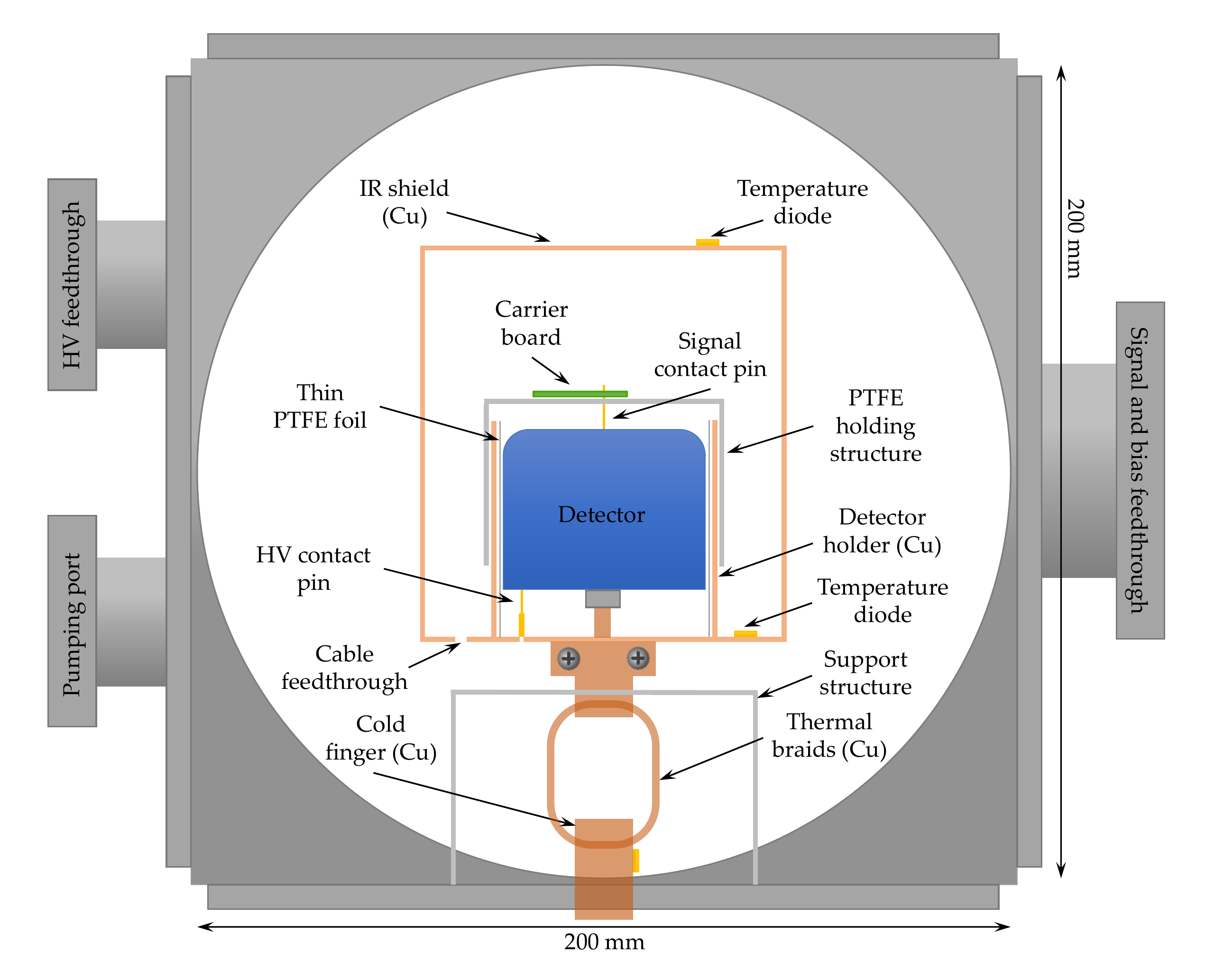}
\caption{Simplified sectional view of the experimental setup used for the investigation of signal readout electronics for LEGEND. For visual clarity, details of the detector holding structure, readout electronics and the cold finger are not shown.}
\label{graph:cube_cryostat}
\end{center}
\end{figure}

An overview of the experimental setup is shown in figure~\ref{graph:cube_cryostat}. The core of the setup is a vacuum cryostat (operated at ${\sim}10^{-7}\,$mbar) that can accomodate a germanium detector as well as the signal readout electronics. During the measurements, the temperature of the detector support structure (measured using a silicon temperature diode at the bottom of the IR shield) was stable at a level of~${\sim}98\,$K.

\begin{figure}[!ht]
\centering\quad
\begin{varwidth}[b]{0.5\linewidth}
\begin{tabular}{ll}
\toprule
Property & Value \\
\midrule
Mass    								&$1.0\,$kg	\\
Inner diameter $a$						&$58.9\,$mm \\
Outer diameter $b$						&$68.9\,$mm \\
Length $c$								&$52.0\,$mm \\
Length $d$ 								&$47.0\,$mm \\
Deadlayer (Ge/Li) $e$					&$1.24\pm0.10\,$mm\\
Capacitance                 			&$1.8\,$pF	\\
Depletion voltage						&$900\,$V	\\
\bottomrule
\end{tabular}
\quad
\end{varwidth}
\begin{minipage}[b]{0.5\linewidth}
\centering
\includegraphics[angle=0,width=0.87\textwidth]{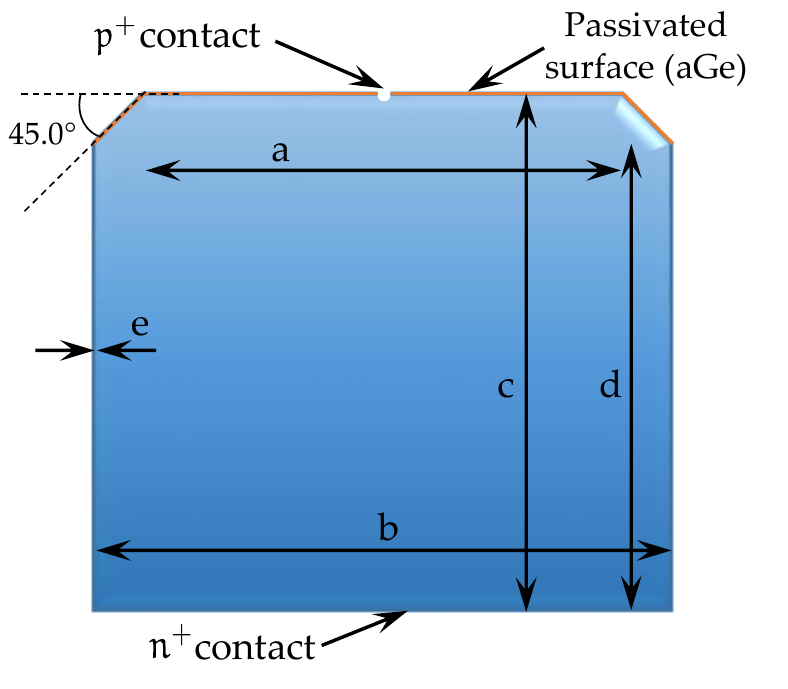}
\vspace{-2.8cm}
\end{minipage}
\caption{Parameters and sketch of the PPC detector used for the CUBE ASIC characterization measurements.}
\label{graph:detector_dimensions}
\end{figure}

To evaluate and characterize the ASIC-based signal readout electronics, a p-type point contact germanium detector was used. Due to their distinct geometry, see figure~\ref{graph:detector_dimensions}, PPC detectors have a very low detector capacitance ($C_{\text{det}}{\sim}1-2\,\text{pF}$ at full depletion) \cite{luke1989}. This not only results in a low energy threshold, but also in a good noise performance. Furthermore, PPC detectors have an excellent ability to discriminate events that deposit their energy at a single location (single-site events, indicator for $0\text{\textnu\textbeta\textbeta}$ decay signal events) from those that deposit their energy at multiple sites (multi-site events, e.g.~Compton-scattered photon, indicator for background events). This is due to the strongly localized weighting potential and the strong electric field in close proximity to the p$^+$ signal readout contact leading to characteristic signal pulse shapes \cite{mertens2019}.
\FloatBarrier
\noindent
The detector used in this work is a natural germanium PPC detector with properties that closely resemble those of the detectors currently operated in the \textsc{Majorana Demonstrator}, see figure~\ref{graph:detector_dimensions}. In our studies, the detector was installed with the point contact facing up in a customized detector mount. It was shielded against IR radiation (emitted mainly by the vacuum cryostat walls) by a thin cylindrical copper hat cooled via liquid nitrogen surrounding the holding structure, see figure~\ref{graph:cube_cryostat}. The n$^+$ electrode of the detector was connected to the high voltage module via a spring-loaded pin located at the detector bottom. To reduce high-frequency voltage fluctuations introduced by the high voltage power supply, an $RC$ low-pass filter ($100\,$M$\text{\textOmega}$, $10\,$nF) was used. Data from the detector were recorded with a Struck SIS$3301$ $14$-bit flash analog-to-digital converter (FADC).
\FloatBarrier
\noindent

\section{Measurement results}\label{ch:results}

\subsection{Leakage current}
\begin{figure}[!h]
\begin{center}
\includegraphics[angle=0,width=0.6\textwidth]{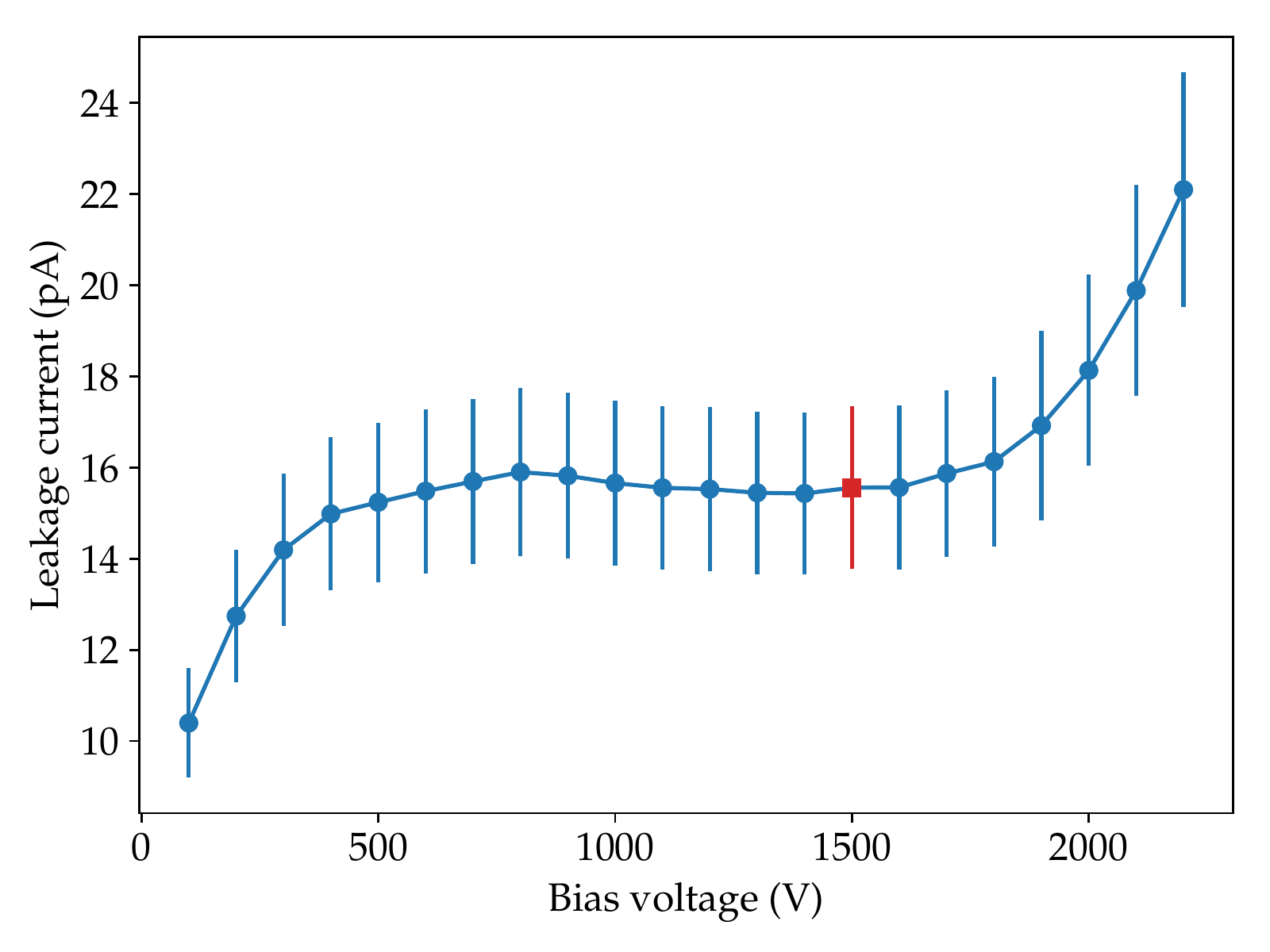}
\caption{Leakage current as a function of the detector bias voltage. Error bars result from the uncertainties of the feedback capacitance, waveform slope and gain. For the measurements discussed in the following, the detector was operated at a bias voltage of $V_{\text{B}}=1500\,$V (indicated by the red measurement point).}
\label{graph:leakage_current}
\end{center}
\end{figure}

The leakage current $I_{\text{leak}}$ has an important impact on the electronic noise, with higher leakage currents resulting in higher noise levels. Hence, dedicated measurements were performed to determine the leakage current of the setup. Several pulsed reset waveforms were acquired with an oscilloscope and the slope of the linearly decreasing waveform ramps (corresponding to the constant collection of holes) was estimated. The leakage current was then calculated using the equation
\begin{align}
I_{\text{leak}}=C_{\text{f}}\cdot\frac{\text{d}V}{\text{d}t}\cdot\frac{1}{G},
\label{eq:leakage_current}
\end{align}
where $V$ denotes the voltage, $t$ the time and $G=4.03\pm0.47$ an additional gain introduced by the biasing board of the ASIC readout. The dependence of the leakage current on the bias voltage is shown in figure~\ref{graph:leakage_current}. For the measurement results presented in the following, the detector was operated at a bias voltage of $V_{\text{B}}=1500\,$V. At this bias voltage, the leakage current was at a reasonably low level of $I_{\text{leak}}\approx15\,$pA. At the time when the energy resolution measurements were carried out, see section~\ref{ch:results_resolution}, the leakage current was stable at a level of $10-20\,$pA.

\subsection{Signal rise time}
\begin{figure}[!h]
\begin{center}
\mbox{
\hspace{-0.5cm}
\subfigure[Signal rise time.]{{\includegraphics[width = 0.48\textwidth]{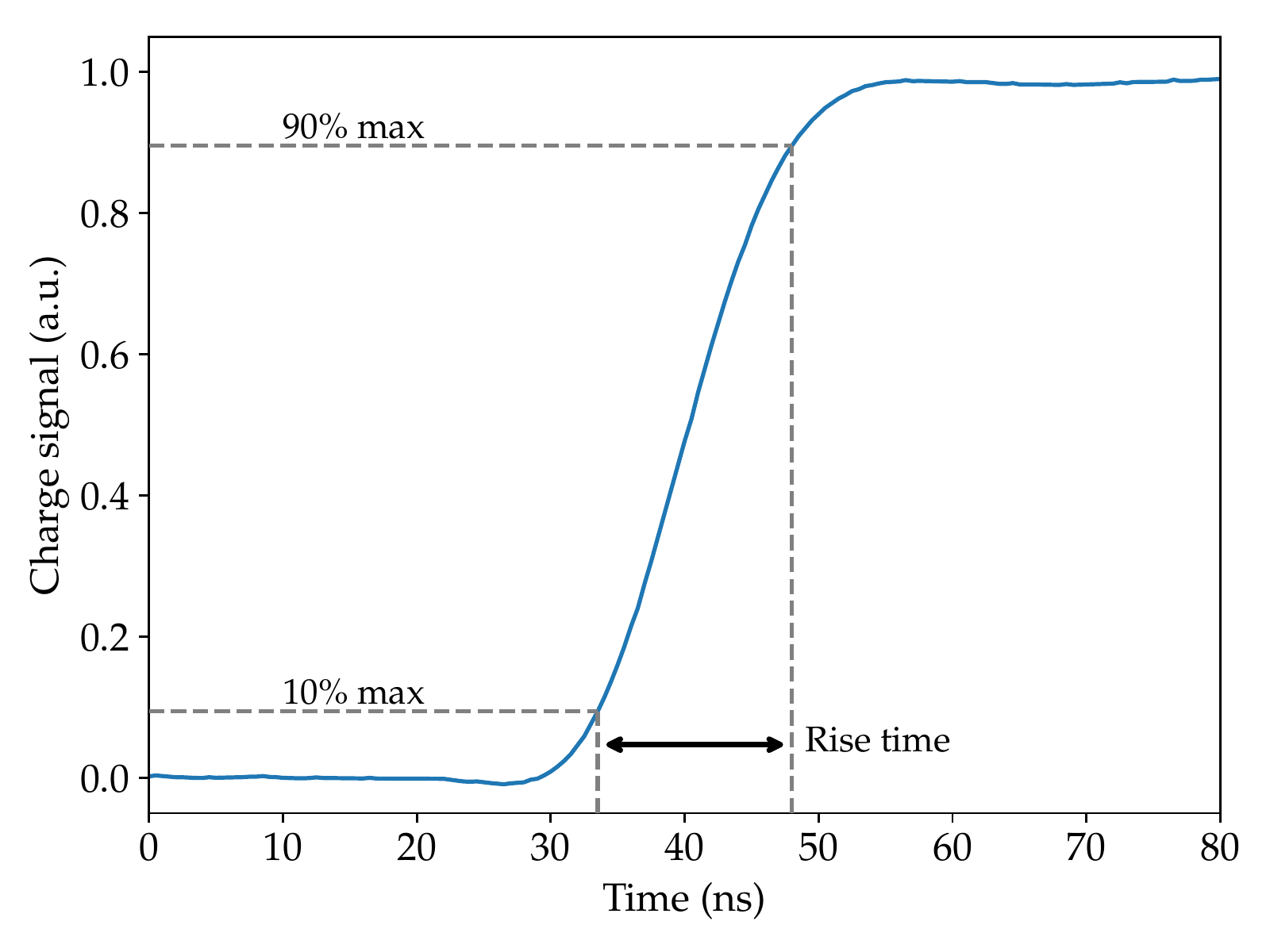}}}\quad
\subfigure[Baseline noise curve.]{{\includegraphics[width = 0.48\textwidth]{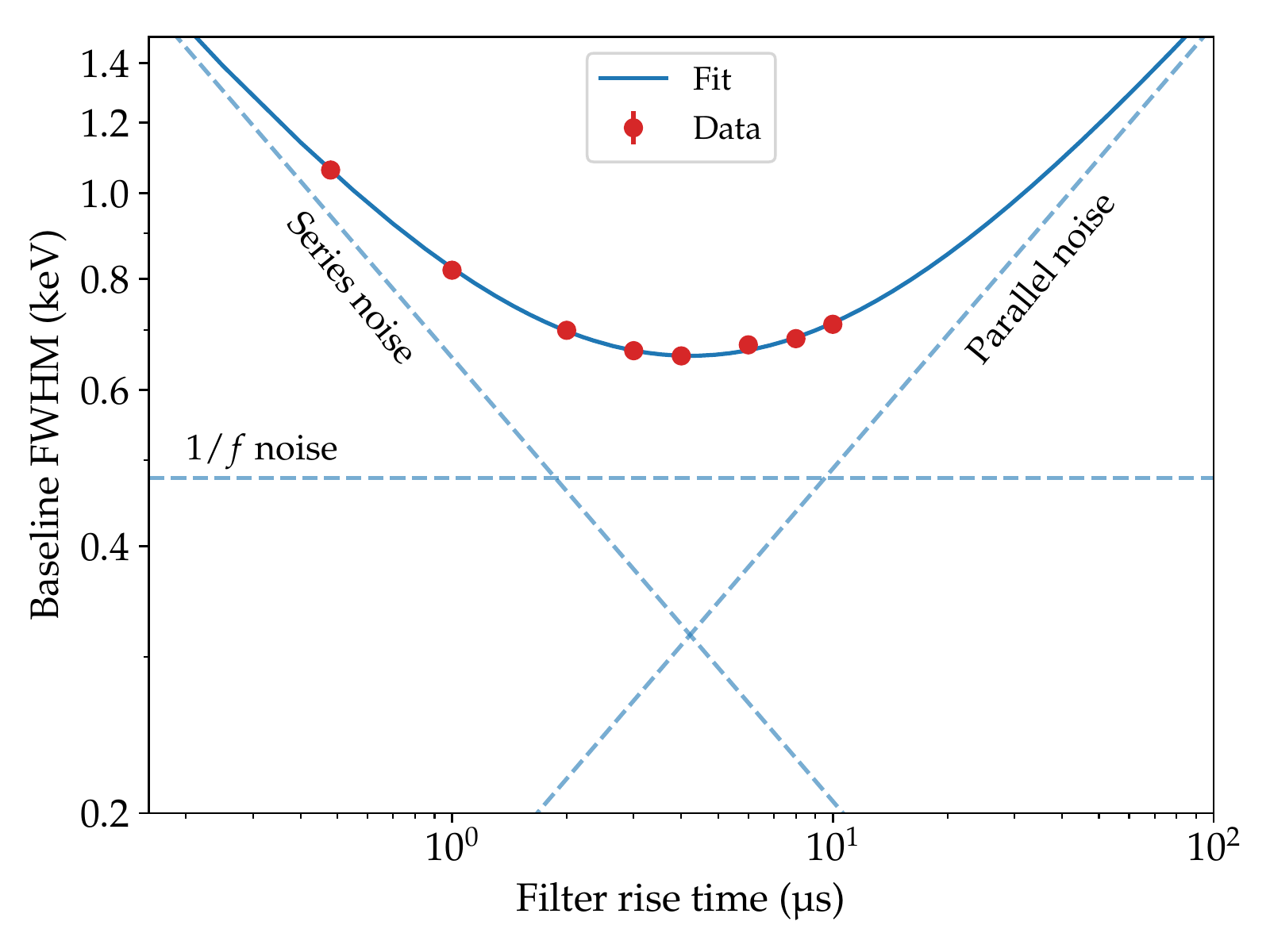}}}
}
\caption{Electronic key parameters of signal readout electronics for $0\text{\textnu\textbeta\textbeta}$ decay searches: signal rise time (a) and electronic noise (b). A signal rise time as low as 15\,ns was measured. The noise performance was investigated in terms of the baseline noise. A minimum baseline noise of 655\,eV FWHM was obtained at a filter rise time of $4\,\text{\textmu s}$. The dashed lines in the plot correspond to the series (down-sloping), parallel (up-sloping) and $1/f$ (horizontal) noise contributions.}
\label{graph:noise_risetime}
\end{center}
\end{figure}

One of the key parameters for a successful application of pulse shape discrimination techniques, discussed in detail in section~\ref{ch:results_psd}, is the signal rise time. Typically, this quantity is defined as the time taken by a signal to change from 10\% to 90\% of the maximum amplitude of the leading edge. For efficient pulse shape discrimination capabilities, the rise time needs to be fast enough such that multi-site events (background events) can be resolved in the time domain (for rise times of several $\text{\textmu s}$, the double peak structure of these events in the current signal smears out). The specification for the first phase of LEGEND foresees rise times faster than 100\,ns ($10\%-90\%$) with potential for improvements in future phases. A dedicated measurement without a detector was carried out to measure the rise time of the CUBE ASIC, see figure~\ref{graph:noise_risetime}\,(a). Rise times as low as 15\,ns were obtained.
\FloatBarrier
\noindent

\FloatBarrier
\noindent

\vspace{-0.3cm}
\subsection{Electronic noise}
Low electronic noise of signal readout electronics is of major importance for optimizing the energy resolution and the detection threshold of the measurement system. Furthermore, along with the signal rise time the electronic noise is a key parameter for efficient pulse shape discrimination capabilities. While low-frequency noise ($\mathcal{O}(\text{kHz})$) mainly influences the energy resolution, high-frequency noise ($\mathcal{O}(\text{MHz})$) has an impact on the PSA performance. The relevant frequency range for the application of pulse shape analysis techniques is given by the necessity of resolving the temporal separation of multi-site event (MSE, see section~\ref{ch:results_psd}) charge clouds with values $150-500\,$ns, which translates into a frequency range of $2.0-6.5\,$MHz~\cite{wagner2017}.

The noise performance of the CUBE ASIC together with the PPC detector was determined in terms of the baseline noise. To this end, a trapezoidal filter with varying filter rise times and a fixed flat top time was applied to the waveform baselines. The obtained baseline noise curve is shown in figure~\ref{graph:noise_risetime}\,(b). At a filter rise time of $4\,\text{\textmu s}$, a minimum baseline noise of 655\,eV FWHM was obtained. Moreover, at the reference filter rise time of $1\,\text{\textmu s}$, we measured a baseline noise of about 820\,eV FWHM (corresponding to 118\,e$^-$~RMS). Keeping in mind the presence of the additional detector and bonding capacitance, this value is in good agreement with the specified preamplifier noise performance. In summary, fast signal rise times and low noise levels make the CUBE ASIC a well-suited device for the application of pulse shape analysis techniques.

\subsection{Energy resolution}\label{ch:results_resolution}
\begin{figure}[!h]
\begin{center}
\vspace{-0.3cm}
\mbox{
\hspace{-0.5cm}
\subfigure[$^{228}$Th energy spectrum.]{{\includegraphics[width = 0.515\textwidth]{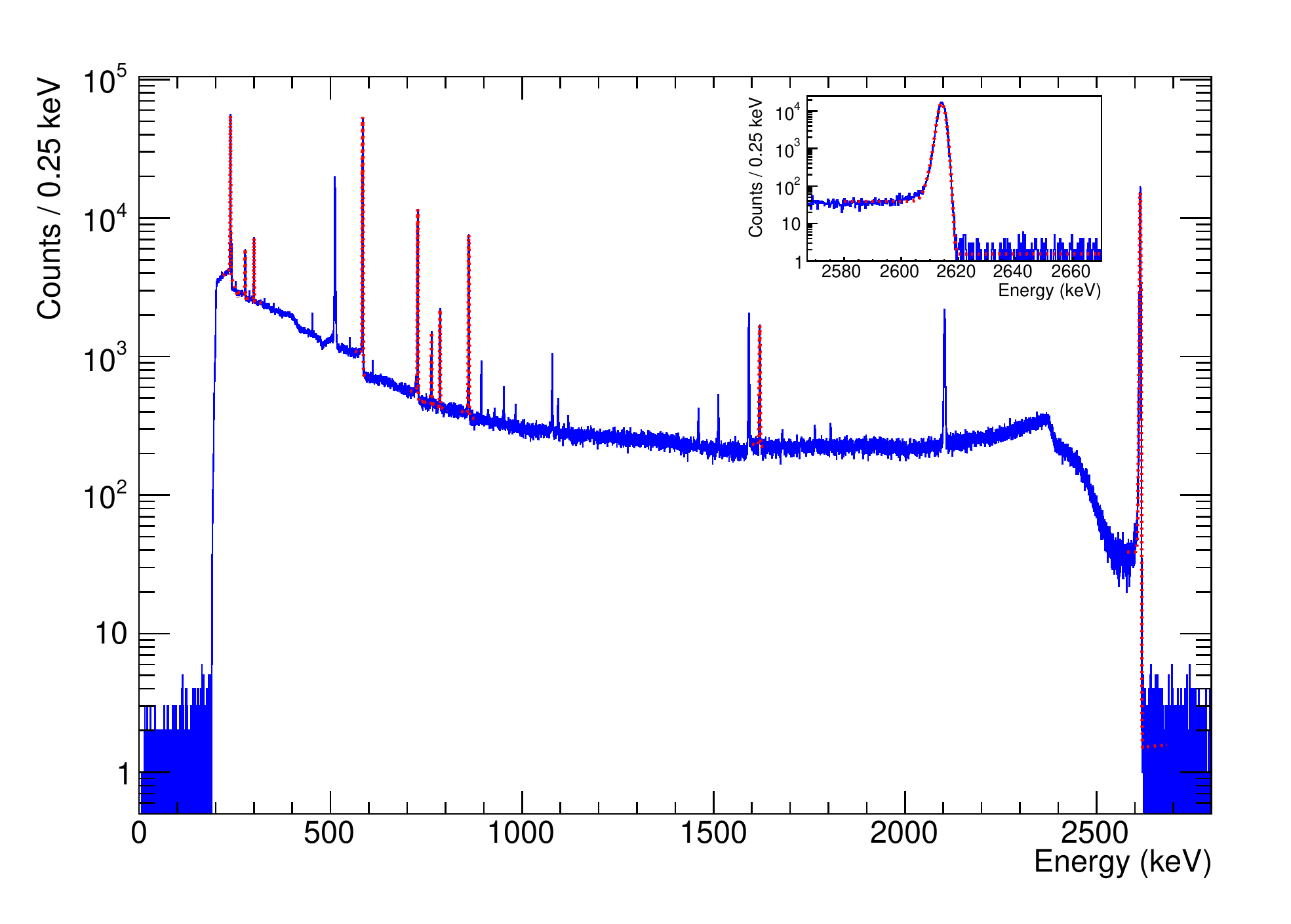}}}
\subfigure[Energy resolution.]{{\includegraphics[width = 0.515\textwidth]{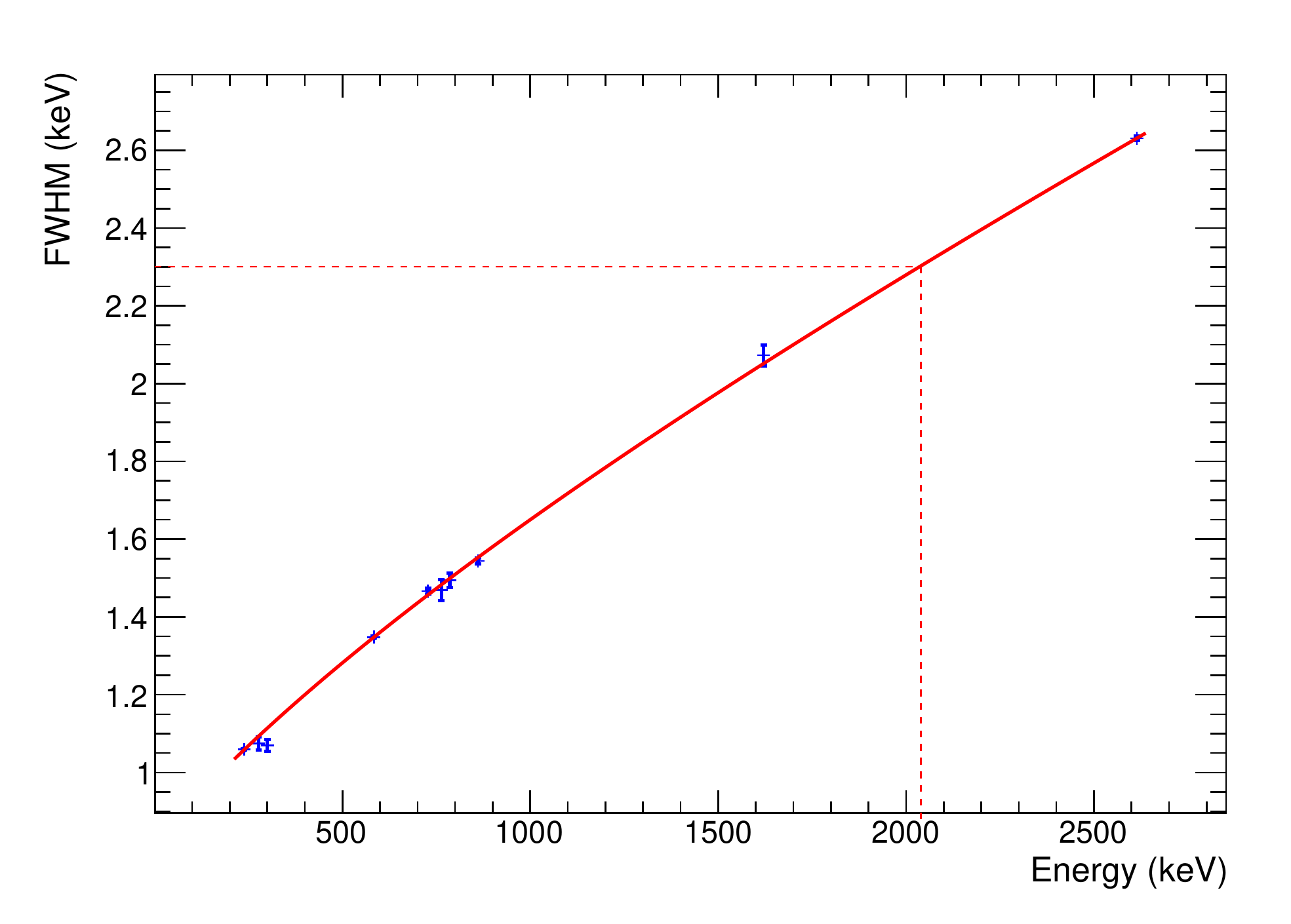}}}
}
\caption{Energy spectrum (a) and resolution curve (b) acquired during a typical $^{228}$Th calibration run. The energy resolution of about 2.3\,keV FWHM in the signal region of interest at $Q_{\text{\textbeta\textbeta}}=2039\,$keV is indicated by the dashed lines. At the 2.6\,MeV $^{208}$Tl gamma line (see inset), an energy resolution of about 2.6\,keV FWHM was obtained. Error bars correspond to the fit uncertainties of the standard deviations.}
\label{graph:energy_resolution}
\end{center}
\end{figure}

The excellent energy resolution of germanium detectors is one of the main advantages of $^{76}$Ge-based $0\text{\textnu\textbeta\textbeta}$ decay searches. The energy resolution is closely related to the noise performance of the signal readout electronics, i.e.~high noise levels directly translate into poor energy resolutions. Therefore, dedicated measurements were carried out to investigate the energy resolution of the PPC detector together with the CUBE ASIC. To this end, the detector was irradiated with a strong, collimated $^{228}$Th calibration source. The source was positioned outside the vacuum cryostat in front of one of the side flanges. The signal rate was on the order of $350\,$counts/s. An example of the energy spectrum measured during a typical $^{228}$Th calibration run is shown in figure~\ref{graph:energy_resolution}\,(a).

For the calibration of the energy scale $E$ and to obtain an estimate for the energy resolution, several known gamma lines in the spectrum were fit using a function \cite{benato2015} of the form
\begin{equation}
\begin{split}
f(E)&=A\exp\left(-\frac{\left(E-\text{\textmu}\right)^2}{2\sigma^2}\right)+B+\frac{C}{2}\text{erfc}\left(\frac{E-\text{\textmu}}{\sqrt{2}\sigma}\right)+D(E-\text{\textmu})\\
&+\frac{F}{2}\exp\left(\frac{E-\text{\textmu}}{\delta}\right)\text{erfc}\left(\frac{E-\text{\textmu}}{\sqrt{2}\sigma}+\frac{\sigma}{\sqrt{2}\delta}\right)
\end{split}
\end{equation}
where $A, B, C$, $D$ and $F$ describe normalization factors, $\text{\textmu}$ the mean and $\sigma$ the standard deviation of a Gaussian distribution and $\delta$ the decay constant of an exponential. The second, third and fourth terms describe the background shape of the energy spectrum underlying the gamma peaks. The last term is an exponentially modified low-energy Gaussian tail used to approximate the peak shape distortion due to incomplete charge collection \cite{alvis2019, benato2015}. The energy resolution $\Delta E$ is typically described in terms of the full width at half maximum (FWHM) of a gamma line at energy $E$. It is computed by numerically extracting the difference of the two half-maximum points of the fit function. The energy resolution as a function of the energy for the $^{228}$Th calibration measurement is shown in figure~\ref{graph:energy_resolution}\,(b). The resolution can be described by the following expression \cite{benato2015, agostini2019-2}:
\begin{align}
\text{FWHM}(E)&=2\sqrt{2\log{2}}\sqrt{\sigma_{\text{ENC}}^2+\sigma_{\text{CP}}^2+\sigma_{\text{CC}}^2}\\
&=2\sqrt{2\log{2}}\sqrt{\frac{\text{\texteta}^2}{e^2}\text{ENC}^2+\text{\texteta} FE + c^2E^2}.\label{eq:fwhm}
\end{align}
Here, $\sigma_{\text{ENC}}$ describes the electronic noise, $\sigma_{\text{CP}}$ statistical fluctuations in the charge production process and $\sigma_{\text{CC}}$ the efficiency of the charge collection process in the detector. Furthermore, $\text{\texteta}$ describes the average energy needed to create an electron-hole pair in germanium, $\text{ENC}$ the equivalent noise charge, $F$ the Fano factor and $c$ a constant. As can be seen from figure~\ref{graph:energy_resolution}\,(b), an excellent energy resolution over a wide energy range was obtained. In the signal region of interest at the $Q_{\text{\textbeta\textbeta}}$-value and at the 2.6\,MeV $^{208}$Tl gamma peak, energy resolutions of about 2.3\,keV~FWHM and 2.6\,keV~FWHM were obtained, respectively. These values match the design specifications of LEGEND-1000, with a targeted energy resolution of 2.5\,keV FWHM at the $Q_{\text{\textbeta\textbeta}}$-value.
\FloatBarrier
\noindent

\subsection{Pulse shape discrimination performance}\label{ch:results_psd}
\begin{figure}[!h]
\begin{center}
\mbox{
\hspace{-0.5cm}
\subfigure[Single-site event (SSE).]{{\includegraphics[width = 0.48\textwidth]{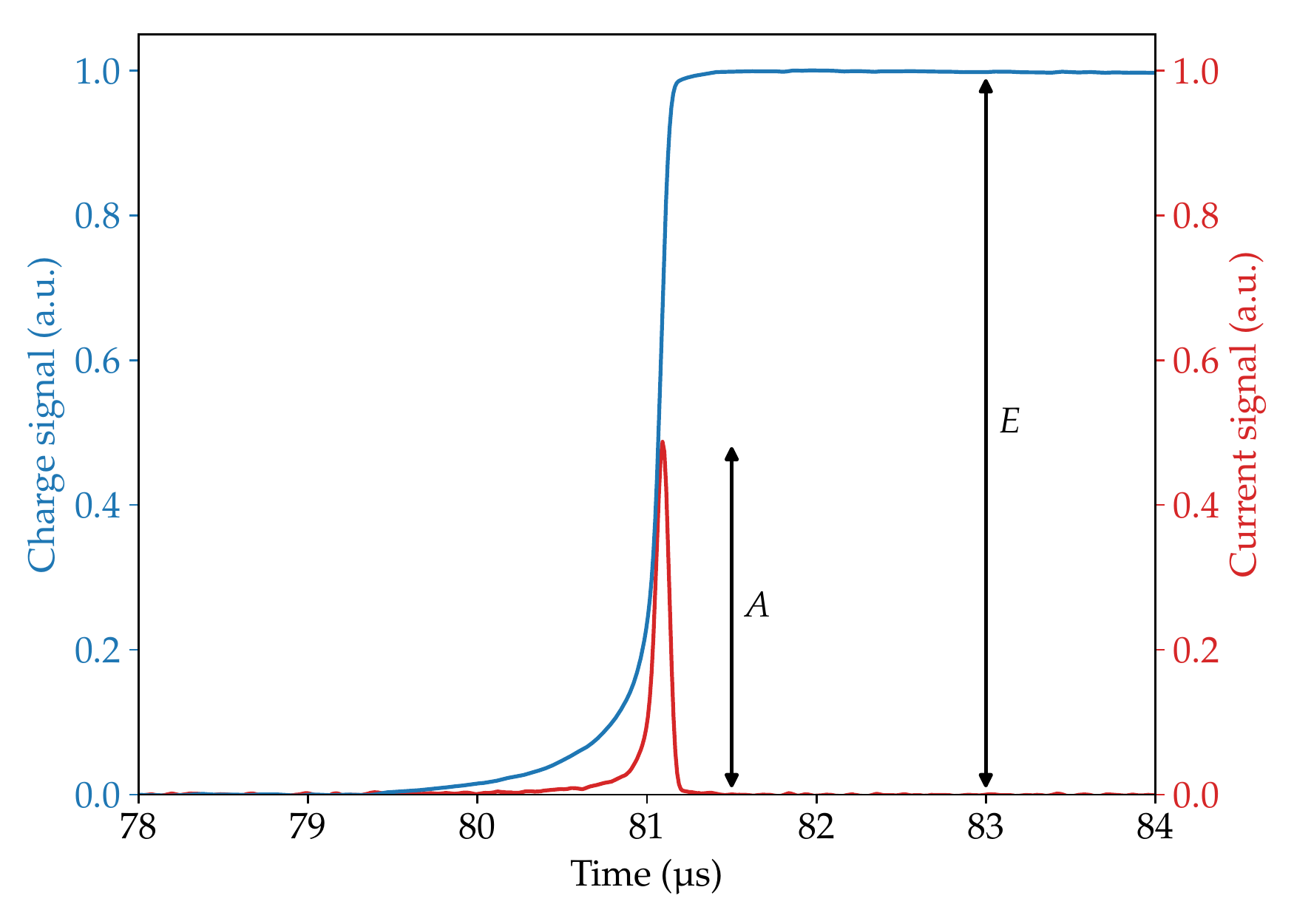}}}\quad
\subfigure[Multi-site event (MSE).]{{\includegraphics[width = 0.48\textwidth]{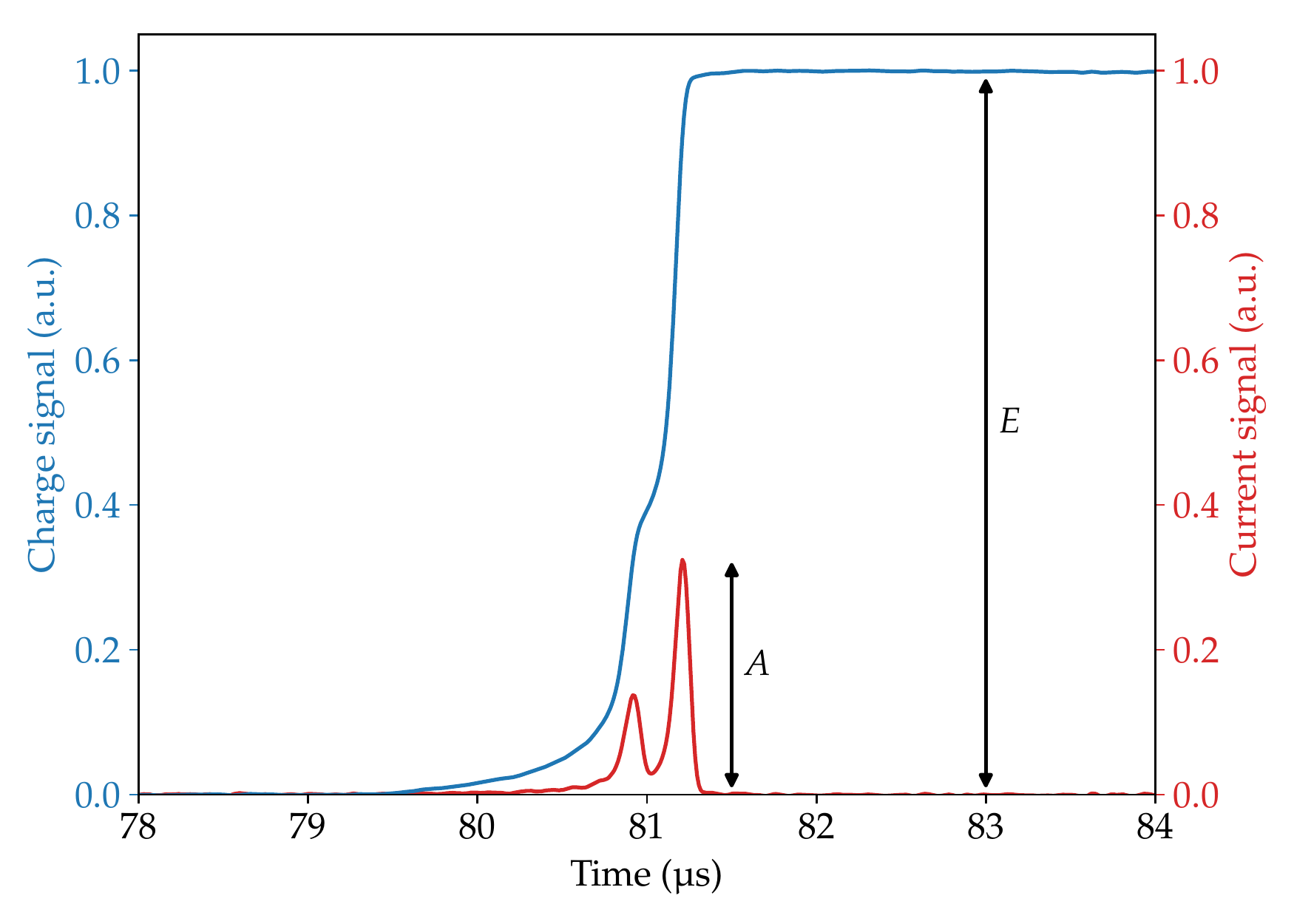}}}
}
\caption{Charge and current signals corresponding to a single-site event (a) and a multi-site event (b) acquired with the CUBE ASIC. The current pulse (red line) corresponds to the time derivative of the charge pulse (blue line). The different signal shapes of a single- and multi-site event can be clearly identified by the different maximal heights $A$ of the current pulses.}
\label{graph:sse_mse_examples}
\end{center}
\end{figure}

In order to fulfill the ultra-low background requirements for $0\text{\textnu\textbeta\textbeta}$ decay searches, it is important to appropriately discriminate background events from signal events. One powerful background rejection method is based on the analysis of the shape of the signal pulses, commonly referred to as pulse shape analysis (PSA) or pulse shape discrimination (PSD). While in the vicinity of the PPC detector's signal readout electrode the weighting potential is strong and highly localized, it is relatively low elsewhere in the active volume. As a consequence, the signal shapes of events with a single energy deposition location in the detector (single-site events, SSE) are almost independent of their point of origin. In contrast, events with multiple energy deposition locations in the detector (multi-site events, MSE) clearly deviate from this shape. An example for a SSE and a MSE acquired using the ASIC-based readout electronics is shown in figure~\ref{graph:sse_mse_examples}. The difference in the signal shape can be seen easily, with two distinct interactions evident in the multi-site event. $0\text{\textnu\textbeta\textbeta}$ decay signal-like events occur at a single location in the germanium crystal (both electrons are stopped within an unresolvable distance of $1\,$mm) and are thus SSE. In contrast, background events (from gamma-ray interactions) usually deposit energy at multiple locations in the detector and are thus MSE.
\FloatBarrier
\noindent
A commonly used discriminative quantity of the signal pulse shape is the ratio of the maximum amplitude of the current pulse $A$ and the amplitude (energy) of the charge pulse $E$: $A/E$, see figure~\ref{graph:sse_mse_examples} \cite{agostini2013}. The $A/E$ distribution of SSE is narrow and only slightly dependent on the energy. In contrast, the $A/E$ distribution of MSE is broad and located at lower values due to the reduced maximum current amplitudes compared to SSE.

In practice, $^{228}$Th is an isotope typically used for defining $A/E$ cuts for the discrimination of signal-like from background events. In the radioactive decay of the isotope to the stable nucleus $^{208}$Pb, high energy gammas with an energy of $2614.5\,$keV are produced. In the detector, they are likely to undergo pair production producing an electron-positron pair. The positron stops and then can form positronium together with another electron. This system is unstable and the particles annihilate each other emitting two $511\,$keV gammas back to back. These gammas can either fully deposit their energy in the detector (full energy peak at $2614.5\,$keV, FEP) or either one or both gammas escape the detector. If only one gamma escapes the detector, an energy of $2614.5\,\text{keV}-511\,\text{keV}=2103.5\,\text{keV}$ (single escape peak, SEP) is deposited in the detector. In contrast, if both gammas escape the active volume, an energy of $2614.5\,\text{keV}-2\cdot511\,\text{keV}=1592.5\,\text{keV}$ (double escape peak, DEP) is deposited in the detector. While the DEP is a good indicator for SSE (energy deposition only by the initial electron), the SEP is used as an indicator for MSE (energy deposition at multiple sites). The extra peak at $1620.5\,$keV close to the DEP can be associated to gamma radiation emitted during the decay of the isotope $^{212}$Bi in the $^{228}$Th decay chain.

In the data analysis, the $A/E$ pulse shape discriminator is tuned such that 90\% of the SSE (signal-like events) in the DEP survive (a detailed description of the procedure can be found in \cite{wagner2017}). The survival efficiency $\text{\textepsilon}$ of the events in the Compton continuum, SEP and FEP can be computed accordingly. Usually, the number of multi-site events in the SEP for PPC detectors can be reduced to below 10\% \cite{alvis2019}.
\begin{figure}[!h]
\begin{center}
\mbox{
\hspace{-0.5cm}
\subfigure[$A/E$ vs $E$ scatter plot.]{{\includegraphics[width = 0.50\textwidth]{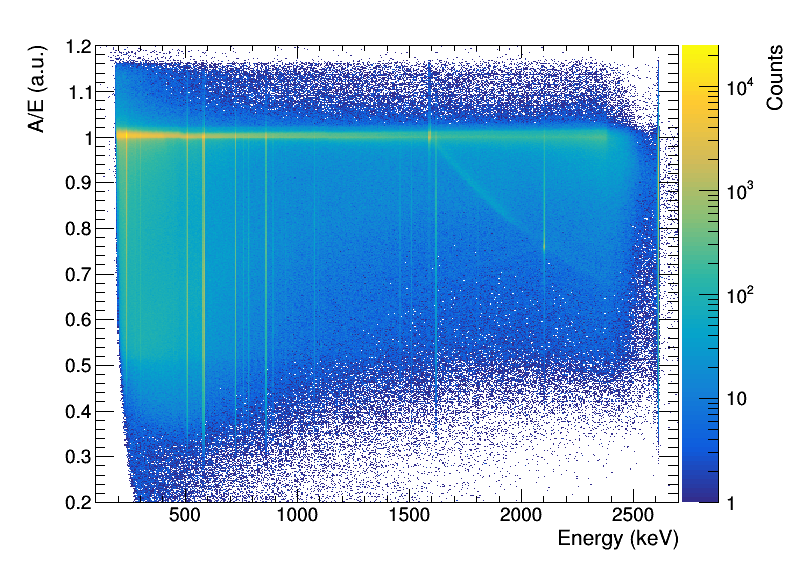}}}\quad
\subfigure[Energy spectra before and after $A/E$ cut.]{{\includegraphics[width = 0.50\textwidth]{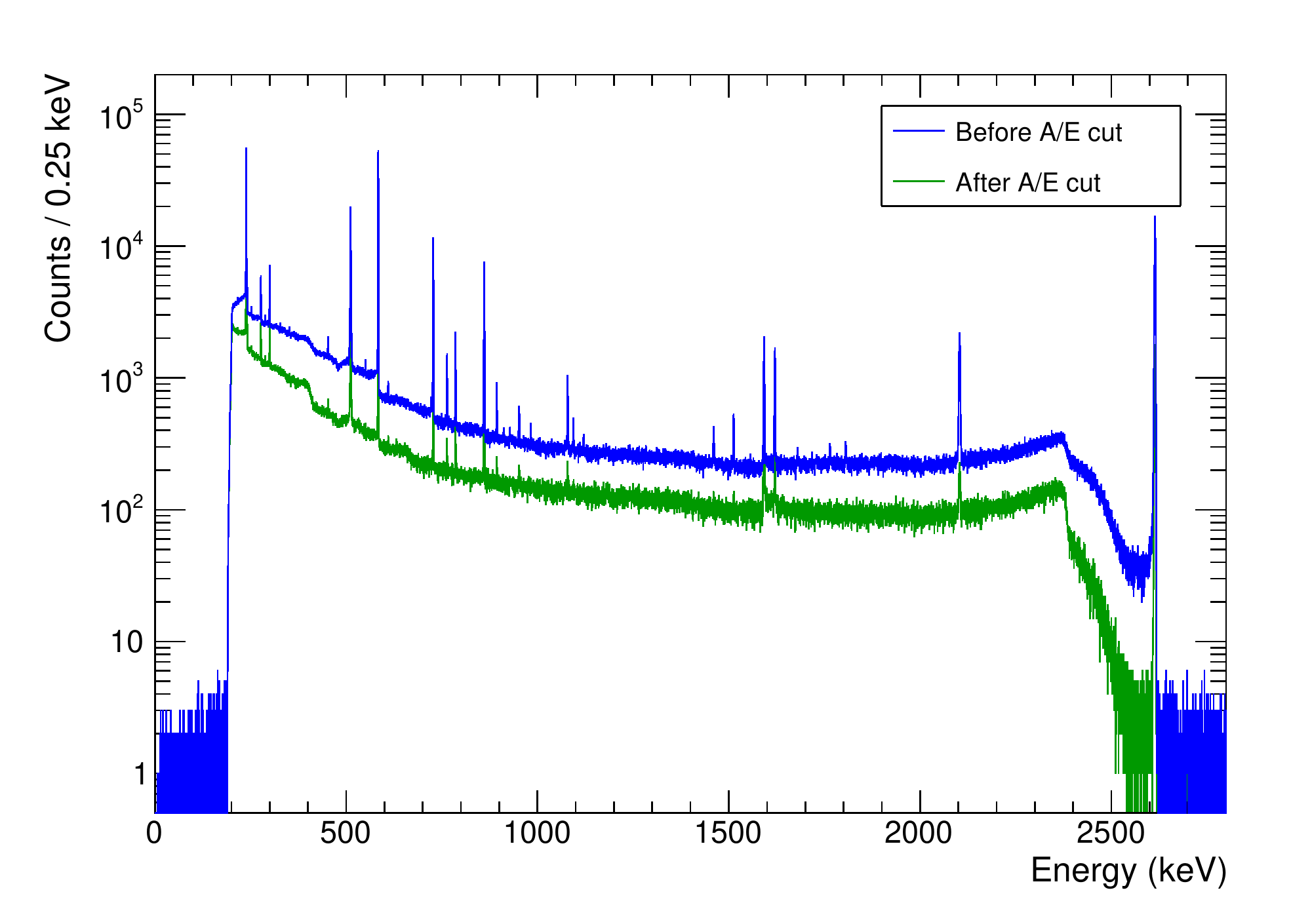}}}
}
\\
\mbox{
\hspace{-0.5cm}
\subfigure[DEP before and after $A/E$ cut.]{{\includegraphics[width = 0.50\textwidth]{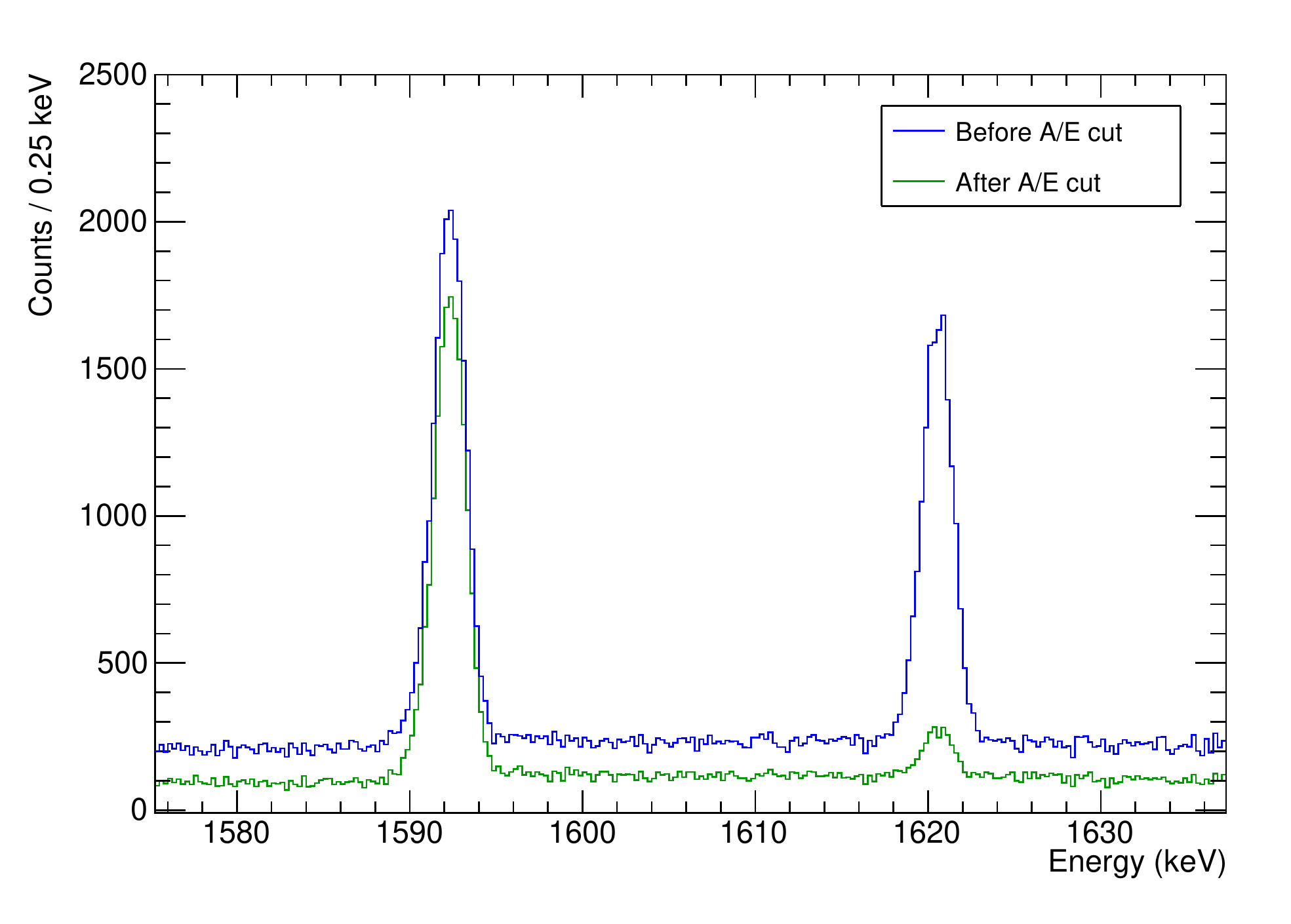}}}\quad
\subfigure[SEP before and after $A/E$ cut.]{{\includegraphics[width = 0.50\textwidth]{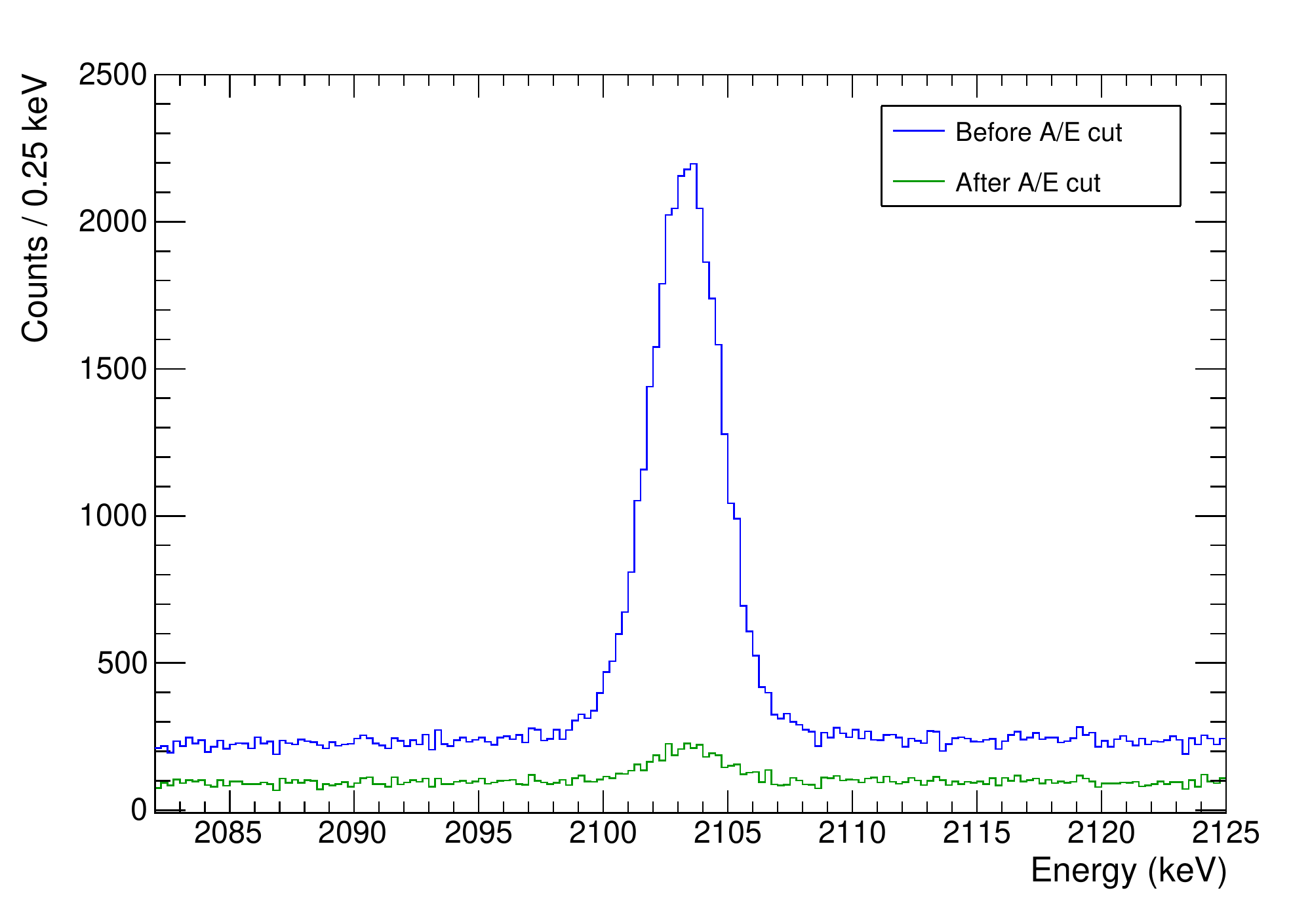}}}
}
\caption{Pulse shape discrimination performance of the PPC detector with the CUBE ASIC: Normalized $A/E$ as a function of the energy for a $^{228}$Th calibration run (a) and energy spectra before and after the $A/E$ pulse shape analysis cut (b). While the cut is tuned such that 90\% of the single-site events (signal-like events) in the DEP survive (c), the number of multi-site events (background events) in the SEP is significantly reduced (d).}
\label{graph:psa_performance}
\end{center}
\end{figure}
\FloatBarrier
\noindent
The PSA performance of the PPC detector together with the CUBE ASIC was validated in a $^{228}$Th calibration measurement. The normalized $A/E$ distribution, corrected for a slight linear energy dependence, is shown in figure~\ref{graph:psa_performance}\,(a). The band corresponding to single-site events at $A/E=1$ can be clearly identified. Events below this band mainly correspond to MSE and can be suppressed by applying a cut to the data, based on the $A/E$ parameter, called the $A/E$ cut. The estimated survival efficiencies $\text{\textepsilon}$ are listed in table~\ref{tab:aoe_efficiencies}. The acceptance of background events (MSE) in the SEP is heavily suppressed, i.e.~only 6.0\% of these events survive. At the same time, the acceptance of events in the signal region of interest is 42\%. The effect of the $A/E$ cut on the energy spectrum is depicted in figures~\ref{graph:psa_performance}\,(b)-(d). The survival efficiencies obtained in this work are in good agreement with the efficiencies obtained with the same detector type in the \textsc{Majorana Demonstrator} experiment \cite{alvis2019, alvis2019-1}.
\begin{table}[!h]
\caption{Pulse shape discrimination performance of the PPC detector with the CUBE ASIC. The estimation of the survival efficiencies $\text{\textepsilon}$ is based on the $A/E$ pulse shape discriminator. The acceptance of events in the double escape peak (DEP, mostly single-site events) is tuned to 90\%. The survival efficiencies of (mostly multi-site) events in the single escape peak (SEP), full energy peak (FEP), as well as in the region of interest at the $Q_{\text{\textbeta\textbeta}}$-value can then be evaluated. The uncertainties of the survival efficiencies correspond to statistical uncertainties.}
\begin{center}
\begin{tabular}{cc}
\bottomrule
Peak&Survival efficiency $\text{\textepsilon}$\\
\midrule
$^{208}$Tl DEP ($1592.5\,$keV)	&$0.900\pm0.007$\\
$^{208}$Tl SEP ($2103.5\,$keV)	&$0.060\pm0.004$\\
$^{208}$Tl FEP ($2614.5\,$keV)	&$0.106\pm0.001$\\
$Q_{\text{\textbeta\textbeta}}$ ($2039\,$keV)	&$0.420\pm0.002$\\
\bottomrule
\end{tabular}
\label{tab:aoe_efficiencies}
\end{center}
\end{table}
\FloatBarrier
\noindent

\subsection{Radiopurity}
A high radiopurity of the components used in a low-background physics experiment like \mbox{LEGEND} is crucial. In order to predict the background rate that would be induced by the CUBE ASIC in future phases of LEGEND, it has been assayed. These assay measurements were performed at the low-background screening facilities at LNGS in Italy and at Jagiellonian University in Poland by means of direct gamma-ray counting, mass spectroscopy and radon emanation techniques, respectively. All assay results are listed in table~\ref{tab:radiopurity_assay}.

First, the contamination of 35.3\,g ASIC material (leftover production material) in the isotopes $^{232}$Th, $^{238}$U as well as $^{40}$K was analyzed via gamma-ray counting. To this end, the radioactivity of the material was measured with a HPGe detector for a time period of about 23.5\,days. In the measurement, only upper limits were obtained for the contaminations. At the same time, the $^{232}$Th and $^{238}$U impurities in the ASIC were investigated via high-resolution inductively-coupled plasma mass spectrometry (ICP-MS). To accomplish this, 17\,mg of sample material were dissolved in HF, HNO$_3$ and HCl. Finally, the radon contamination of the ASIC was investigated. A cryogenic radon detector at Jagiellonian University was used to determine the contaminations in 25\,g sample material in both radon isotopes, i.e.~$^{222}$Rn and the short-lived $^{220}$Rn. Just as for the other radiopurity measurements, only upper limits were obtained, see table~\ref{tab:radiopurity_assay}.

Based on the gamma counting measurement results and the efficiencies obtained with \textsc{Geant4} simulations of the LEGEND-200 detector array (used as an approximation for the LEGEND-1000 detector array) \cite{abgrall2018}, the background contribution of the CUBE ASIC can be estimated. For the determination of the efficiencies $p_{\text{ROI}}$ (counts/decay/keV) in the signal region of interest, a detector anti-coincidence cut was applied. The simulation results are listed in table~\ref{tab:background_indices}. Usually, the radioactive background is expressed in terms of the background index BI ($\text{counts}/(\text{keV}\cdot\text{kg}\cdot\text{yr})$):
\begin{align}
\text{BI}=\frac{p_{\text{ROI}}\cdot m_{\text{ASIC}}\cdot a}{m_{\text{DET}}}\label{eq:background_index}.
\end{align}
Here, $m_{\text{ASIC}}$ denotes the mass of the radioactive ASIC material ($m_{\text{ASIC}}=500\cdot0.33\,$mg) and $m_{\text{DET}}$ the total detector mass in the LEGEND-1000 detector array ($m_{\text{DET}}=500\cdot2\,$kg) assuming 500\,channels. Furthermore, $a$ describes the specific activity (Bq/kg). The background indices of the $^{232}$Th and the $^{238}$U radionuclides are listed in table~\ref{tab:background_indices}. They were calculated using the specific activities of the $^{228}$Th and $^{226}$Ra contributions of the gamma counting measurements, see table~\ref{tab:radiopurity_assay}. The contributions of the other radionuclides were neglected since they are not relevant for the background in the region of interest. Even though the upper limits of the assay are comparably high (mainly determined by the sensitivity of the measurement method), the summed background index of the $^{232}$Th and the $^{238}$U contributions matches the design specification of LEGEND-1000, with an overall background goal of $1\cdot10^{-5}\,\text{counts}/(\text{keV}\cdot\text{kg}\cdot\text{yr})$.
\begin{table}[!h]
\caption{Results of the CUBE ASIC radiopurity assay conducted at LNGS and Jagiellonian University by means of direct gamma counting, ICP-MS and radon emanation techniques. All values are upper limits.}
\begin{center}
\begin{tabular}{lllll}
\bottomrule
\multirow{2}{*}{Method} &\multicolumn{2}{c}{\multirow{2}{*}{Radionuclide}} & \multicolumn{2}{c}{Purity}\\
& & &\multicolumn{1}{c}{mBq/kg} &\multicolumn{1}{c}{g/g}\cr
\midrule
\multirow{6}{*}{$\text{\textgamma}$ counting}
& $^{232}$Th:
&$^{228}$Ra			&$<4.9$	&$<1.2\cdot10^{-9}$	\\
& &$^{228}$Th	    &$<4.1$	&$<1.0\cdot10^{-9}$	\\
& $^{238}$U:
&$^{234}$Th			&$<24$	&$<1.9\cdot10^{-9}$	\\
& &$^{234\text{m}}$Pa&$<200$	&$<1.6\cdot10^{-8}$	\\
& &$^{226}$Ra		&$<3.5$	&$<2.8\cdot10^{-10}$\\
& $^{40}$K 		&	&$<52$	&$<1.7\cdot10^{-6}$	\\
\midrule
\multirow{2}{*}{ICP-MS}
& $^{232}$Th	&	&		&$<2.0\cdot10^{-9}$ \\
& $^{238}$U		&	&		&$<1.0\cdot10^{-9}$	\\
\midrule
\multirow{2}{*}{Rn emanation}
& $^{220}$Rn	&	&$<0.8$	& \\
& $^{222}$Rn	&	&$<0.9$	& \\
\bottomrule
\end{tabular}
\label{tab:radiopurity_assay}
\end{center}
\end{table}
\FloatBarrier
\noindent

\begin{table}[!h]
\caption{Efficiencies ($p_{\text{ROI}}$) and background indices (BI) of the CUBE ASIC for the radionuclides $^{232}$Th and $^{238}$U. Upper limits correspond to 90\% confidence level. The efficiency values are based on \textsc{Geant4} simulations of the LEGEND-200 detector array and are used as an approximation for the LEGEND-1000 detector array \cite{abgrall2018}.}
\begin{center}
\begin{tabular}{lcc}
\bottomrule
Radionuclide &$p_{\text{ROI}}$ $(\text{counts}/(\text{decay}\cdot\text{keV}))$ &BI $(\text{counts}/(\text{keV}\cdot\text{kg}\cdot\text{yr}))$\\
\midrule
$^{232}$Th	&$1.92\cdot10^{-5}$	&$<4.1\cdot10^{-7}$ \\
$^{238}$U	&$9.62\cdot10^{-6}$	&$<1.8\cdot10^{-7}$ \\
\bottomrule
\end{tabular}
\label{tab:background_indices}
\end{center}
\end{table}
\FloatBarrier
\noindent

\section{Conclusions and outlook}\label{ch:conclusion}
Signal readout electronics based on application-specific integrated circuit (ASIC) technology are ideally suited for low-background $0\text{\textnu\textbeta\textbeta}$ decay experiments like LEGEND. While not being inferior to discrete readout systems, ASIC technology could allow for a lower electronic noise, a lower per-channel power consumption, and a higher per-channel radiopurity (less mass close to the detectors).

In this work, we carried out a detailed investigation of the performance of a commercially available ASIC, the XGLab CUBE charge sensitive amplifier (CSA). The ASIC was operated together with a p-type point contact high-purity germanium detector. Dedicated measurements were carried out to investigate key electronic parameters. The studies reveal that 1) an excellent energy resolution over a wide energy range, 2) very fast signal rise times and 3) low noise levels can be obtained with an ASIC CSA. These parameters are important for the effective application of pulse shape analysis techniques (PSA) for the discrimination of signal events from background events. The PSA performance of the ASIC-based readout system (acceptance of background events in a $^{228}$Th calibration measurement: ${\sim}6\%$) was found to be comparable to the performance reported by the \textsc{Majorana Demonstrator} experiment \cite{alvis2019, alvis2019-1}. Finally, the radiopurity of the CUBE preamplifier was analyzed at the low-background screening facilities at LNGS in Italy and at Jagiellonian University in Poland by means of various assay techniques. In all measurements, upper limits for the radioactive contaminations were obtained that are compatible with the background goal of LEGEND-1000.

The results presented in this work are very promising for a potential application of ASIC technology in LEGEND-1000. However, the investigated CUBE ASIC is not ideally suited for the final application in LEGEND: The bypass capacitors required for filtering the noise of the ASIC power supplies increase the amount of radioactive material close to the detectors and are probably not compatible with the required radiopurity levels. In future ASIC developments, this can be alleviated by using a single internally filtered power supply or ultrapure bypass capacitors (e.g.~silicon capacitors). 
To further reduce radioactive contamination, the reset mechanism needs to be integrated into the chip. A continuous reset mode with exponentially decaying pulses is foreseen. Moreover, a differential output is required to reduce the noise associated with driving signals over long transmission lines from the detectors to the data acquisition system. A dedicated ASIC fulfilling these requirements is currently being developed at Lawrence Berkeley National Laboratory (LBNL).

In conclusion, the performed measurements offer valuable information about the operation of a large-scale germanium detector together with an ASIC-based signal readout system. The results presented in this study provide vital information for the design of future ASIC-based readout systems for the final phase of LEGEND.

\acknowledgments
The authors would like to thank P.~Barton, B.~Lehnert and A.~Poon of Lawrence Berkeley National Laboratory for useful discussions throughout the project and the \textsc{Majorana Demonstrator} collaboration for loaning the germanium detector.

This work was supported by the Max Planck Society, the Technical University of Munich and the DFG Collaborative Research Center "Neutrinos and Dark Matter in Astro- and Particle Physics" (SFB\,1258). F.~Edzards gratefully acknowledges support by the German Academic Scholarship Foundation (Studienstiftung des deutschen Volkes), M.~Willers gratefully acknowledges support by the Alexander von Humboldt Foundation, and S.~Mertens gratefully acknowledges support by the MPRG at TUM program.

This material is based upon work supported by the U.S.~Department of Energy, Office of Science, Office of Nuclear Physics under Award Numbers DE-FG02-97ER41041, DE-FG02-97ER41033, DE-AC05-00OR22725, as well as Federal Prime Agreement DE-AC02-05CH11231. This material is based upon work supported by the National Science Foundation under Grant No.~NSF OISE 1743790 and 1812409.


\end{document}